\pdfoutput=1
\documentclass[preprint,11pt]{elsarticle}

\usepackage[margin=0.8in]{geometry} 

\usepackage{soul}

\usepackage{indentfirst}

\usepackage{color, colortbl}  
\usepackage{xcolor} 
\definecolor{Mycolor1}{HTML}{f90b9b}   
\definecolor{Mycolor2}{HTML}{0042FF}   
\definecolor{LightCyan}{rgb}{0.88,1,1}

\usepackage[normalem]{ulem}

\newcommand{\stkout}[1]{\ifmmode\text{\sout{\ensuremath{#1}}}\else\sout{#1}\fi}

\usepackage{url}
\usepackage{breakurl}
\usepackage[breaklinks]{hyperref}

\usepackage[utf8]{inputenc}
\biboptions{numbers,sort&compress}  

\usepackage{graphicx}

\usepackage{arydshln}  

\usepackage{amssymb}

\usepackage{amsthm} 

\usepackage{amsmath}   

\usepackage{times}
\usepackage{sectsty}
\usepackage{amsfonts}   

\usepackage{stackengine}
\usepackage{caption}
\allsectionsfont{\normalsize} 

\makeatletter
\def\ps@pprintTitle{%
 \let\@oddhead\@empty
 \let\@evenhead\@empty
 \def\@oddfoot{}%
 \let\@evenfoot\@oddfoot}
\makeatother

\begin{document}

\begin{frontmatter}


\title{Epidemic dynamics on metapopulation networks with node2vec mobility}

\author[1]{Lingqi Meng}
\author[1,2,3]{Naoki Masuda\corref{cor1}%
}
\ead{naokimas@buffalo.edu}

\cortext[cor1]{Corresponding author}
\address[1]{Department of Mathematics, State University of New York, Buffalo, NY 14260-2900, USA}
\address[2]{Computational and Data-Enabled Science and Engineering Program, State University of New York, Buffalo, NY 14260-5030, USA}
\address[3]{Faculty of Science and Engineering, Waseda University, Tokyo 169-8555, Japan}

\begin{abstract}

Metapopulation models have been a powerful tool for both theorizing and simulating epidemic dynamics. In a metapopulation model, one considers a network composed of subpopulations and their pairwise connections, and individuals are assumed to migrate from one subpopulation to another obeying a given mobility rule. While how different mobility rules affect epidemic dynamics in metapopulation models has been studied, there have been relatively few efforts on comparison of the effects of simple (i.e., unbiased) random walks and more complex mobility rules. Here we study a susceptible-infectious-susceptible (SIS) dynamics in a metapopulation model, in which individuals obey a parametric second-order random-walk mobility rule called the node2vec. We map the second-order mobility rule of the node2vec to a first-order random walk in a network whose each node is a directed edge connecting a pair of subpopulations and then derive the epidemic threshold. For various networks, we find that the epidemic threshold is large (therefore, epidemic spreading tends to be suppressed) when the individuals infrequently backtrack or infrequently visit the common neighbors of the currently visited and the last visited subpopulations than when the individuals obey the simple random walk. The amount of change in the epidemic threshold induced by the node2vec mobility is in general not as large as, but is sometimes comparable with, the one induced by the change in the diffusion rate for individuals.

\end{abstract}


\end{frontmatter}

\section{Introduction}

One reason for the scientific community to study contact networks to great extents is their relevance to infectious diseases \cite{keeling2005networks, danon2011networks, masuda2013predicting, pellis2015eight, pastor2015epidemic, kiss2017mathematics}. Even in just the last two decades, we have witnessed several global health threats due to infectious diseases, such as SARS pandemic from 2002 to 2003, influenza H1N1 that broke out in 2009, West Africa Ebola outbreak from 2013 to 2016, and the pandemic of COVID-19 that is impacting the entire globe as of 2021. 

In the case of sexually transmitted infections, partnerships are relatively stable over time. In contrast, mobility of individuals causes contact networks to vary over time and affects dynamics of the aforementioned infectious diseases and many others. Metapopulation models provide a mathematically tractable description of epidemic processes under mobility of individual agents in networks \cite{hethcote1978immunization, may1984spatial, lloyd1996spatial, grenfell1997meta, grenfell1998cities, hufnagel2004forecast, colizza2006role, colizza2007reaction, colizza2008epidemic, balcan2009multiscale, masuda2010effects, balcan2011phase, vespignani2012modelling, nicolaides2012metric, tizzoni2015scaling, gomez2018critical, soriano2018spreading, soriano2020vector}. They take advantage of the merits of both the fully mixed property of individuals within each subpopulation (also called patch) and the specific structure of connectivity between the subpopulations. A metapopulation model assumes that a population of individuals is distributed over subpopulations, which correspond to the geographical locations, such as local gathering places, cities, or counties. On the microscopic scale, the individuals are assumed to be fully mixed within each subpopulation. An infectious individual infects each susceptible individual in the same subpopulation with the same rate/probability. This assumption is practical in the absence of detailed data on the structure of interactions among the individuals within each subpopulation. On the macroscopic scale, the individuals traverse edges in the network to travel from one subpopulation to another according to a mobility rule.

Mobility patterns impact epidemic spreading, perhaps as considerably as the network structure. The simplest mobility rule is to assume that each individual moves from the currently visited subpopulation to one of its adjacent subpopulations with the same probability or with probability proportional to the weight of the edge connecting the two subpopulations. Most of early researches of metapopulation models assumed this mobility rule, which one often refers to as the simple random walk in the case of unweighted networks (i.e., no weight on edges). Then, one often lumps together the subpopulations having the same degree (i.e., number of edges that the subpopulation has) to facilitate analytical calculations \cite{colizza2007reaction, colizza2007invasion, colizza2008epidemic, balcan2011phase, vespignani2012modelling}. However, mobility patterns for both human and animal individuals relevant to epidemic spreading are considered to be more complex than is described by the simple random walk, which has prompted studies of different mobility rules. For example, the propensity to move may depend on the degree of the subpopulation that the individual currently visits and the number of the individuals in the subpopulation \cite{colizza2008epidemic}. Furthermore, empirical mobility patterns may be better approximated by non-simple random walks \cite{belik2011natural, balcan2011phase, balcan2012invasion, poletto2013human, rosvall2014memory, scholtes2014causality, matamalas2016assessing} or recurrent mobility patterns \cite{balcan2011phase, balcan2012invasion, gomez2018critical, soriano2018spreading, granell2018epidemic, soriano2020impact, feng2020epidemic}. Other extensions of metapopulation network models include multilayer ones, in which individuals having different mobility patterns are assigned to different network layers \cite{xuan2013reaction, wang2014epidemic, soriano2018spreading}.

The effect of the diffusion rate of the individuals on the extent of epidemic spread is known for metapopulation models. Perhaps counterintuitively, a larger diffusion rate yields a larger epidemic threshold, which implies that epidemic spreading is less likely to occur when individuals diffuse at a faster rate \cite{masuda2010effects, gomez2018critical, soriano2018spreading}. An intuitive explanation of this result is that infectious individuals arising in subpopulations containing relatively many individuals more easily spill over to other subpopulations with fewer individuals if the diffusion rate is higher. Effects of higher-order mobility rules on epidemic spreading in metapopulation models have also been examined \cite{matamalas2016assessing}. In contrast, how mobility rules compare with each other in terms of epidemic spreading has been underexplored. In the present study, we study the effect of the mobility rule modeled by a second-order random walk called the node2vec on epidemic spreading in metapopulation networks. The node2vec is a second-order Markovian random walk first proposed for improving the performance of data mining tasks such as classification of nodes and link prediction \cite{grover2016node2vec}. With the node2vec, one can tune the propensity that the individuals backtrack, which roughly corresponds to the frequency of recurrent mobility, and the weight of the local versus global search of the network. These two properties each correspond to the two parameters of the node2vec random walk. We note that the node2vec includes the simple random walk and non-backtracking random walk as special cases. There have been some theoretical results on the node2vec random walks. Qiu et al. showed that one can factorize a matrix related to the stationary distribution and transition probability of the node2vec as the length of random walks tends to infinity \cite{qiu2018network}. Furthermore, we previously showed that a node2vec random walker diffuses faster than a simple random walker in general when it infrequently travels backwards or visits the common nodes of the currently visited and the last visited nodes \cite{meng2020analysis}. In the present study, we derive the epidemic threshold for the susceptible-infectious-susceptible (SIS) model when individuals obey the node2vec mobility rule. Then, we analyze the epidemic threshold for synthetic and empirical networks of subpopulations. We find that for various networks with a heterogeneous degree distribution, the epidemic threshold is large when individuals tend to avoid backtracking and visiting subpopulations that are an immediate neighbor of the presently visited and last visited subpopulations.

\section{Methods}
\subsection{Metapopulation model with the node2vec mobility rule}
\label{Metapopulation model and node2vec mobility rule}

We examine the continuous-time SIS model on the metapopulation network when individuals move from one subpopulation to another obeying the node2vec random walks. We consider a connected, undirected, and unweighted network with $N$ subpopulations, $M$ edges, and $N\rho$ individuals. Therefore, there are $\rho$ individuals per subpopulation on average. The generalization of the following formulation to directed and weighted networks is straightforward. A subpopulation is a container of individuals and represents a habitat, a college dormitory, an urban area, or a province, for example. Each individual is in either the susceptible or infectious state at any given time. Each infectious individual independently infects each susceptible individual in the same subpopulation at rate $\beta$, which we refer to as the infection rate. This implementation of the infection process for the metapopulation model is known as the type-I reaction \cite{colizza2007reaction, pastor2015epidemic}. Each infectious individual independently recovers to transit to the susceptible state at rate $\mu$, which we refer to as the recovery rate. 

\begin{figure}[!t]
  \centering
  \includegraphics[width=0.99\textwidth]{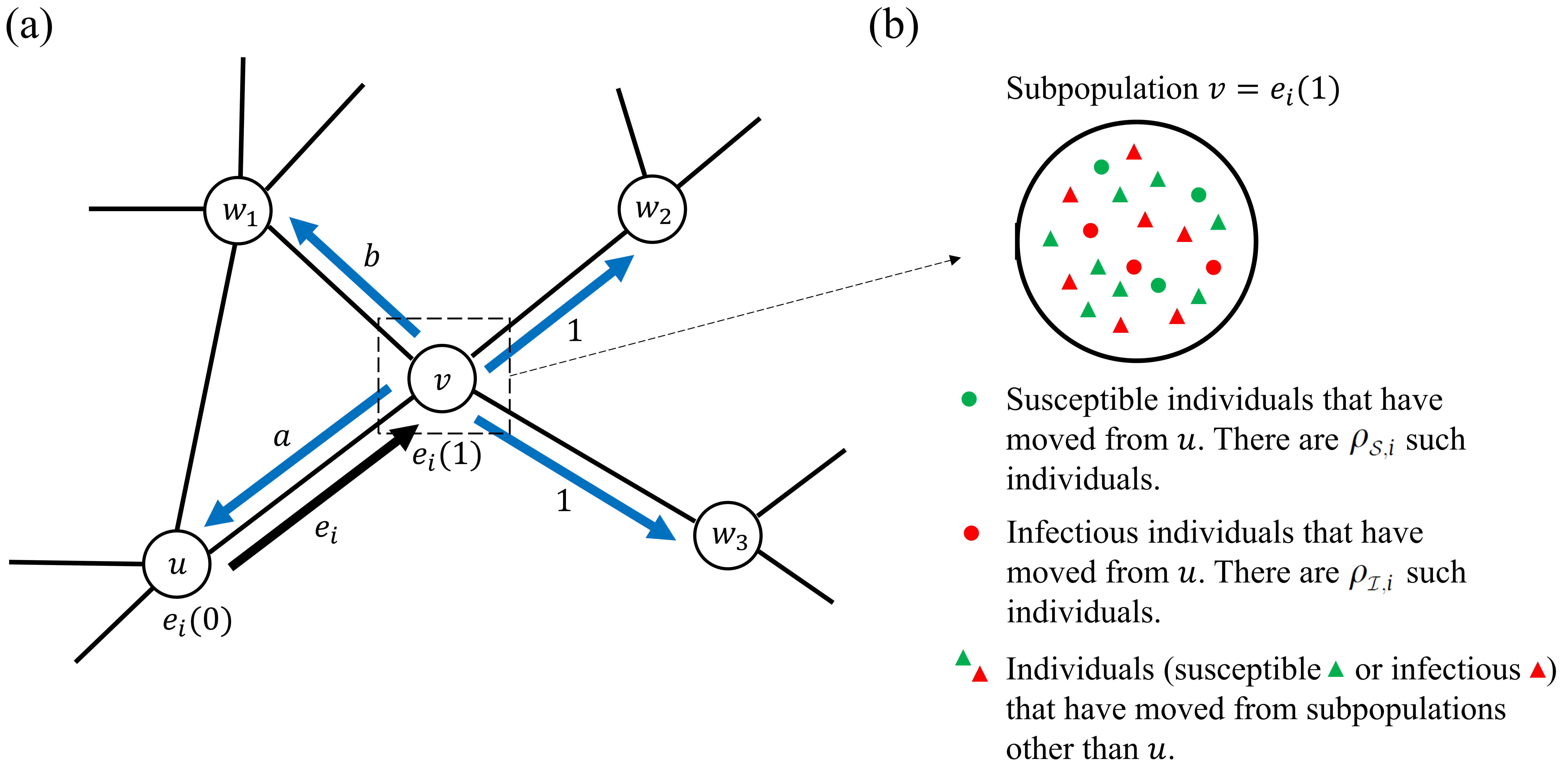}
  \caption{Schematic of the SIS model on metapopulation model networks with the node2vec mobility. (a) Node2vec mobility rule. Node $v$ represents the currently visited subpopulation. Node $u$ represents the subpopulation that the individual visited just before moving to $v$ by traversing directed edge $e_i$ (i.e., by traversing undirected edge $(u, v)$ in the direction from $u$ to $v$). The transition probability from $v$ to one of the four neighbors of $v$ in the next move is given by $a/(a+b+2), b/(a+b+2)$, or $1/(a+b+2)$ in this example. (b) Composition of individuals in a subpopulation.}\label{fig1}
\end{figure}

The individuals also independently perform a continuous-time node2vec random walk on the network of subpopulations as follows. For infinitesimal time $\Delta t$, an individual leaves the currently visited subpopulation, denoted by $v$, with probability $D_\mathcal{S}\Delta t$ or $D_\mathcal{I}\Delta t$ depending on whether the individual is susceptible or infectious, respectively. When an individual leaves $v$, it moves to one of the neighboring subpopulations of $v$. The individual moves back to the subpopulation where the individual was located just before arriving in $v$, which we denote by $u$, with the probability proportional to $a$. The individual moves to a subpopulation that is adjacent to both $v$ and $u$ with the probability proportional to $b$. The individual moves to a different subpopulation with the probability proportional to $1$. For example, in Fig. \ref{fig1}(a), the individual in its next move backtracks to $u$ with probability $a/(a+b+2)$, moves to $w_1$, which is adjacent to $v$ and $u$, with probability $b/(a+b+2)$, or moves to $w_2$ or $w_3$ with probability $1/(a+b+2)$ each. Note that the node2vec mobility reduces to the simple random walk if $a=b=1$. 

\subsection{Stochastic numerical simulations}

We apply the method used in Ref. \cite{colizza2007reaction} to simulate the stochastic SIS dynamics. We set $\rho = 50$. In each simulation, one individual that is selected uniformly at random is initially infectious, and the other $N\rho - 1$ individuals are initially susceptible. We use the rejection method with time steps of length $\Delta t = 10^{-4}$. Specifically, we repeat the following three steps. First, each infectious individual recovers to become susceptible with probability $\mu \Delta t$. Second, each susceptible individual in subpopulation $i$ becomes infectious with probability $1 - (1-\beta \Delta t)^{\mathcal{I}_i}$, where $\mathcal{I}_i$ is the number of infectious individuals in subpopulation $i$. Third, each susceptible and infectious individual independently leaves its current subpopulation with probability $D_\mathcal{S} \Delta t$ and $D_\mathcal{I} \Delta t$, respectively. If this event happens, then the individual migrates to a neighboring subpopulation according to the node2vec mobility rule.

In each simulation, for each pair $(a, b)$ and infection rate $\beta$, we run the simulation for $3\times 10^6$ steps such that the total simulated time is $300$, and then calculate the fraction of infectious individuals in the equilibrium. We calculate the average and standard deviation of the fraction of infectious individuals on the basis of $100$ simulations for each parameter set.

\subsection{Synthetic networks}
\label{Synthetic networks}

In this section, we explain the synthetic networks used in our analysis. For each synthetic network model, except the extended ring network, we set the number of nodes, $N$, to $100$ and repeat generating the network until we obtain a connected network.

In the Erdős–Rényi (ER) model, we place $M$ edges selected uniformly at random between $N(N-1)/2$ pairs of nodes. We set $M=300$ such that the mean degree, $\langle k\rangle$, is $6$. 

In the Barabási–Albert (BA) model, we sequentially add new nodes each with $m=3$ edges that connect to existing nodes according to the linear preferential attachment rule \cite{barabasi1999emergence}. We started the growth process from $N_0=3$ isolated nodes. We connected the first added node to each of the $N_0$ isolated nodes. The degree distribution $p(k)$ approximately obeys $p(k)\propto k^{-3}$, where $\propto$ represents ``proportional to'', in the limit of $N\to \infty$. With $N_0=3$ and $m=3$, there are $291$ edges, which implies $\langle k\rangle = 5.82 \approx 6$. 

The random clustered graph is specified as follows \cite{miller2009percolation, newman2009random}. For each node $i$, parameter $\tilde{t}_i$ represents the number of triangles in which node $i$ participates, and $\tilde{s}_i$ represents the number of edges other than those belonging to the triangles. One draws $N$ random vectors $(\tilde{t}_i, \tilde{s}_i)$, where $i=1,\ldots, N$, from a distribution $\mathcal{F}(\tilde{t}, \tilde{s})$. To form a network, $\sum_{i=1}^N \tilde{t}_i$ must be a multiple of $3$ such that there are $\sum_{i=1}^N \tilde{t}_i/3$ triangles in the network, and $\sum_{i=1}^N \tilde{s}_i$ must be an even number such that no half-edge connected to a node is left unconnected to another node. The generated network may have self-loops or multiple edges. We remove the self-loops and duplicated multiple edges from the generated network. We let $\mathcal{F}(\tilde{t}, \tilde{s})$ be a doubly Poisson degree distribution \cite{newman2009random}, i.e.,
\begin{align}
    \mathcal{F}(\tilde{t}, \tilde{s})=e^{-\mu}\frac{\mu^{\tilde{t}}}{\tilde{t}!}e^{-\nu}\frac{\nu^{\tilde{s}}}{\tilde{s}!},
\end{align}
where the parameter $\mu$ and $\nu$ are the average number of the triangles per node and the average number of edges which do not form a triangle per node, respectively. We set $\mu=2.5$ and $\nu=1$. With self-loops and multiple edges, the expected degree $\langle k \rangle = 2\mu+\nu=6$. After removing self-loops and duplicated multiple edges from a generated network, there are $309$ edges in the final network, which implies $\langle k \rangle=6.18$.

The power-law cluster graph generates a network with degree distribution $p(k)\propto k^{-3}$ with tunable clustering, i.e., density of triangles \cite{holme2002growing}. This model shares the initialization and preferential attachment rule with the BA model. However, when a new node $v$ with $m$ edges joins the existing network, we only carry out the preferential attachment step for the first new edge to connect $v$ to an existing node $u$. For each of the other $m-1$ edges from $v$, we carry out a triad formation step with probability $p$ or the preferential attachment step with probability $1-p$. In a triad formation step, we add an edge from $v$ to a uniformly randomly chosen neighbor of $u$. If $u$ is one of the $N_0$ nodes that initially exist, all the neighbors of $u$ may be already adjacent to $v$ when one attempts to add an edge between $v$ and a neighbor of $u$ using the triad formation step. Then, no more triangles involving both $v$ and $u$ can be formed. In this case, we carry out the preferential attachment even if we have selected the trial formation step with probability $p$. We set $m=3$, $N_0=3$, and $p=0.5$.

The geographical threshold graph model places $N$ nodes independently and uniformly at random in a bounded subset of the $d$-dimensional Euclidean space \cite{masuda2005geographical}. Then, we assign to each node $v$ a weight $w_v$. One joins each pair of nodes, $v$ and $v^\prime$, by an edge if and only if
\begin{align}
    (w_v+w_{v^\prime})h(r)\geqslant \theta.
\end{align}
We do not assume the periodic boundary condition. We let $d=2$ and the bounded subset be a unit square. We set $h(r)=r^{-2}, \theta=80$, and drew the node weights from the exponential distribution with rate parameter $\lambda=1$ independently for the different nodes. The final network has $303$ edges such that $\langle k \rangle =6.06$.

The Lancichinetti–Fortunato–Radicchi (LFR) model generates networks with community structure \cite{lancichinetti2008benchmark}. The degree is designed to obey a power-law distribution with power-law exponent $\gamma$, and the size of the community obeys a power-law distribution with power-law exponent $\kappa$. The model also requires the maximal degree $k_{\mathrm{max}}$ and mean degree $\langle k \rangle$ as input. The mixing parameter $\overline{\mu} \in (0,1)$ specifies the fraction of edges that connect different communities. A small value of $\overline{\mu}$ leads to strong community structure. We set $\gamma=3$, $\kappa=2$, $\langle k \rangle = 6$, $k_{\mathrm{max}}=100$, and $\overline{\mu}=0.1$.

\begin{figure}[!h]
  \centering
  \includegraphics[width=0.47\textwidth]{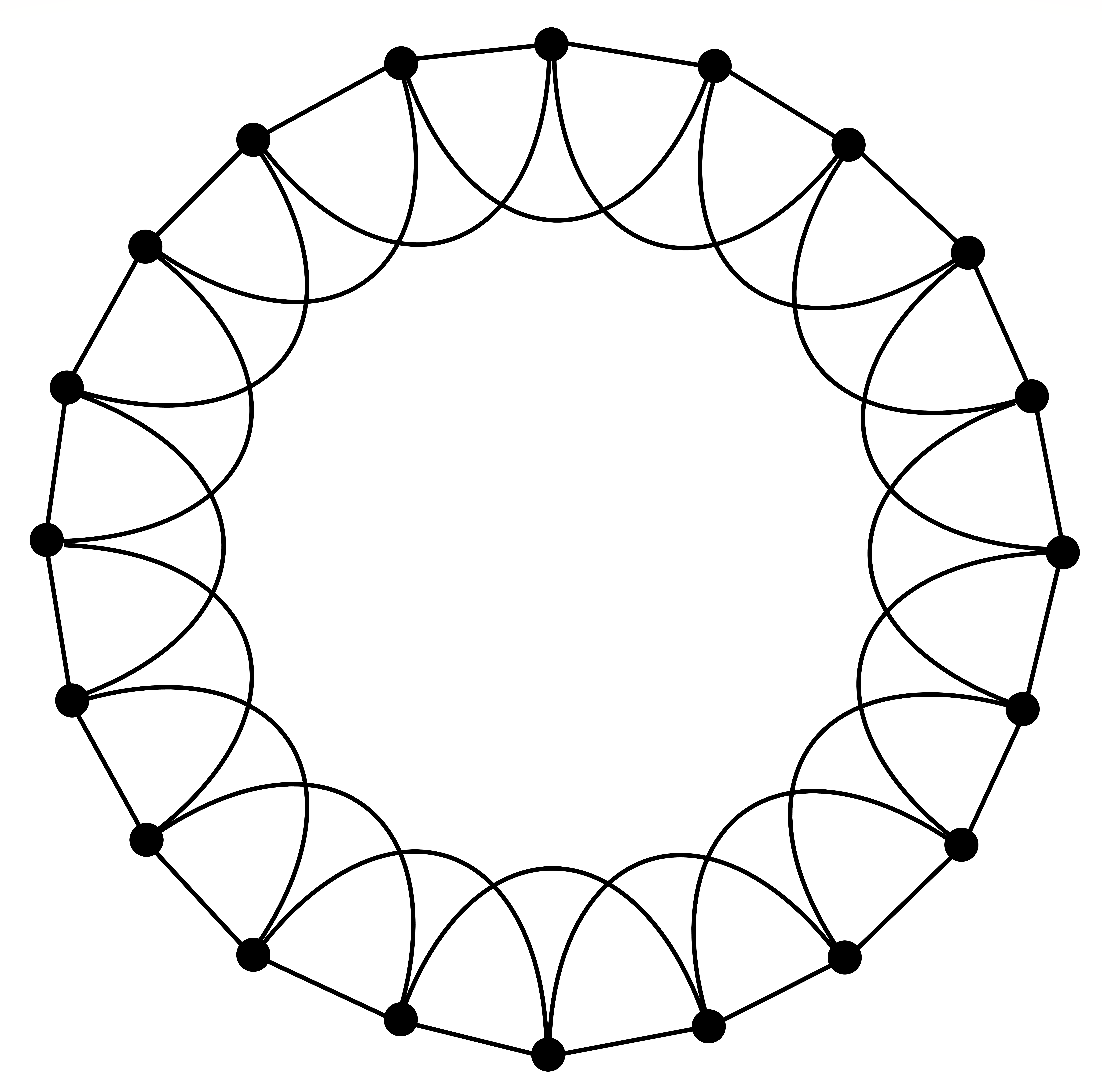}
  \caption{Extended ring network with $N=20$ nodes. }\label{figextend}
  \vspace*{-7pt}
\end{figure}

The extended ring network, an example of which is shown in Fig. \ref{figextend}, is a network with a homogeneous degree distribution. Each node is connected to two neighbors on each side of the ring.

\subsection{Empirical networks}
\label{Empirical networks}

We use the following two empirical networks. The US airport network has $N=332$ airports and $2126$ edges \cite{pajek2006}, and it has a heterogeneous degree distribution. We ignore the weight of edges in the data set. Each subpopulation represents an airport. Two airports are connected if there exists a direct commercial flight between them. 

In the Manizales network, each subpopulation represents a location that individuals visit during working days \cite{lotero2016rich}. Two subpopulations are connected if at least one individual travels between the two locations. The authors of Ref. \cite{lotero2016rich} categorized the individuals in the data set by their socio-economic statuses, from the lowest-income to the wealthiest, into six classes, such that the data set forms a six-layer network. We use the sixth (i.e., wealthiest) layer, which has $45$ subpopulations and $194$ edges. We ignore the weight of edges. 

\section{Results}

\subsection{Master equations}

To calculate the epidemic threshold of the SIS model on metapopulation networks under the node2vec mobility, we start by considering a pair of directed edges in the opposite directions corresponding to each undirected edge. Let $E=\{e_1, \ldots, e_{2M} \}$ be the set of directed edges. We denote each directed edge by $e_i=(e_i(0), e_i(1))\in E$, where $e_i(0)$ and $e_i(1)$ are the source and target node of the directed edge, respectively. One can formulate the node2vec mobility rule, which is a second-order mobility rule in the network of subpopulations, as a first-order Markov chain on a network of the $2M$ directed edges \cite{rosvall2014memory, scholtes2014causality, meng2020analysis}. The transition-probability matrix $T$ of the first-order Markov chain in the network of the directed edges is given by
\begin{align}
	T_{ij}\propto
    \begin{cases} 
      a & \mathrm{if} \ e_i(1)=e_j(0)\ \mathrm{and}\ e_i(0)=e_j(1), \\
      b & \mathrm{if} \ e_i(1)=e_j(0)\ \mathrm{and}\ (e_i(0),e_j(1))\in E, \\
      1 & \mathrm{if} \ e_i(1)=e_j(0)\ \mathrm{and}\ (e_i(0),e_j(1))\not \in E, \\
      0 & \mathrm{otherwise}.
   \end{cases}
   \label{eq edge2}
\end{align}
The normalization is given by $\displaystyle \sum_{j=1}^{2M} T_{ij}=1$ for $i=1, \ldots, 2M$.

The total infection rate in subpopulation $e_i(1)$ at time $t$ is equal to $\displaystyle \beta \sum_{k; e_k(1)=e_i(1)} \rho_{\mathcal{S},k}(t) \sum_{k; e_k(1)=e_i(1)}\rho_{\mathcal{I},k}(t)$, where $\rho_{\mathcal{S},i}(t)$ is the number of susceptible individuals that are located in subpopulation $e_i(1)$ at time $t$ and were located in subpopulation $e_i(0)$ just before arriving at $e_i(1)$. A similar definition applies to $\rho_{\mathcal{I},i}(t)$ (see Fig. \ref{fig1}(b)). We omit $t$ in the following text. The master equations that describe the dynamics of the number of susceptible and infectious individuals in the different subpopulations are given by
\begin{align}
    \frac{d\rho_{\mathcal{S},i}}{dt}= - \beta \rho_{\mathcal{S},i} \sum_{k; e_k(1)=e_i(1)} \rho_{\mathcal{I},k} + \mu \rho_{\mathcal{I},i} - D_\mathcal{S} \rho_{\mathcal{S},i} + D_\mathcal{S} \sum_{k=1}^{2M} \rho_{\mathcal{S}, k} T_{ki}
    \label{eq1}
\end{align}
and
\begin{align}
    \frac{d\rho_{\mathcal{I},i}}{dt}=\phantom{-} \beta \rho_{\mathcal{S},i} \sum_{k; e_k(1)=e_i(1)} \rho_{\mathcal{I},k} -\mu \rho_{\mathcal{I},i} - D_\mathcal{I} \rho_{\mathcal{I},i} + D_\mathcal{I} \sum_{k=1}^{2M} \rho_{\mathcal{I}, k} T_{ki},
    \label{eq2}
\end{align}
where $i=1, \ldots, 2M$. The first term on the right-hand side of Eq. (\ref{eq1}) represents the infection events within subpopulation $e_i(1)$. The sum $\displaystyle \sum_{k; e_k(1)=e_i(1)} \rho_{\mathcal{I},k}$ is equal to the number of infectious individuals at subpopulation $e_i(1)$ regardless of the subpopulation where they were located just before moving to $e_i(1)$. The second term represents the recovery. The third term is the rate of diffusion out of $e_i(1)$ for the susceptible individuals. The last term is the rate of diffusion into $e_i(1)$. Note that $\rho_{\mathcal{S}, k} T_{ki}$ is the number of susceptible individuals that are located in subpopulation $e_k(1)$ and move to subpopulation $e_i(1)$. Note that $T_{ki} = 0$ if $e_k(1)\neq e_i(0)$, which excludes the possibility that individuals move to a subpopulation that is not adjacent to the currently visited subpopulation. A similar interpretation applies to Eq. (\ref{eq2}). 

\begin{figure}[!h]
  \centering
  \includegraphics[width=0.63\textwidth]{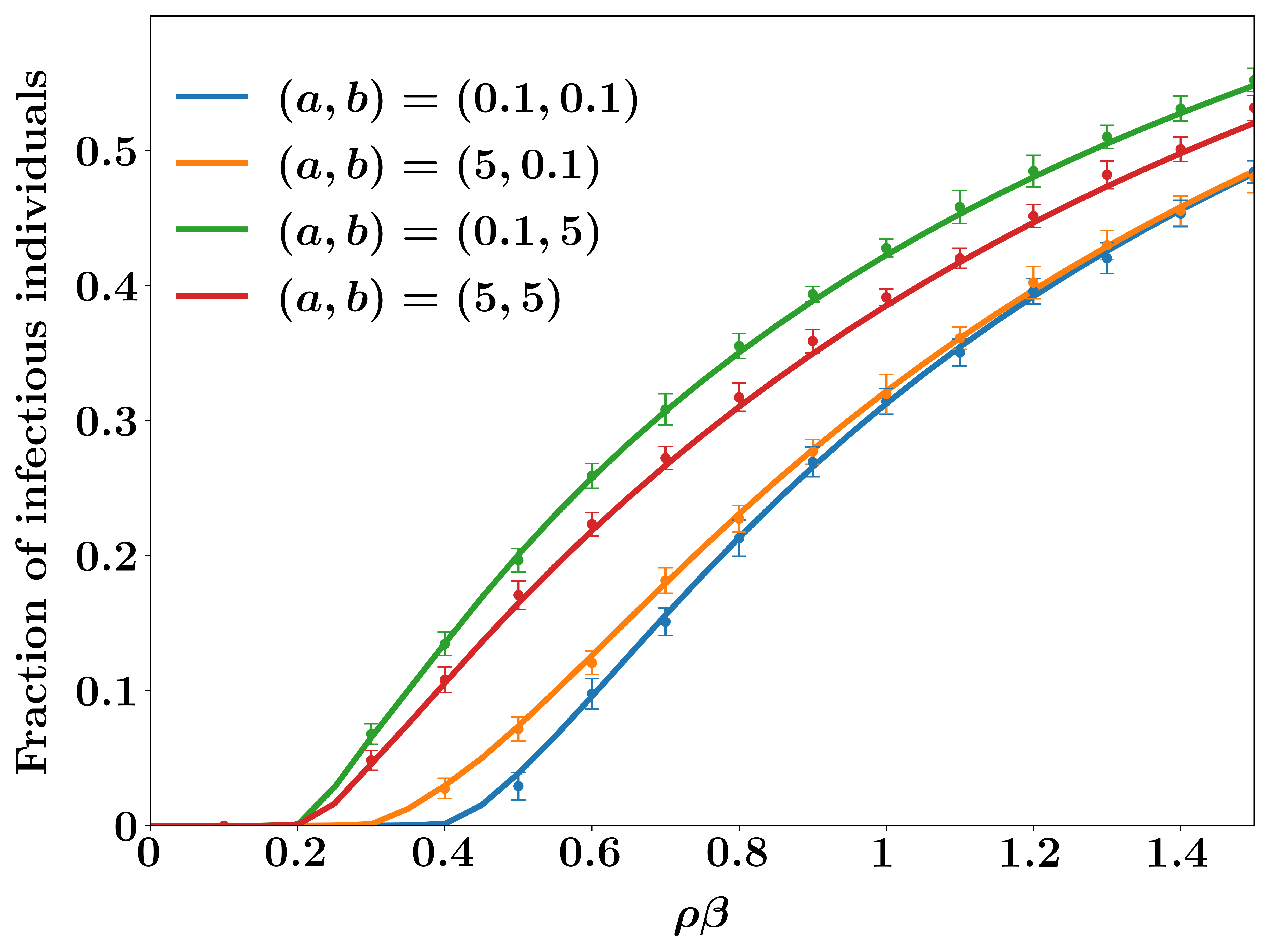}
  \caption{Fraction of infectious individuals in the stationarity as a function of $\rho \beta$ on a network with $N=100$ nodes generated by the BA model. We set $\rho=50$ and consider four sets of $(a,b)$ values. Each circle and error bar represent the average and standard deviation, respectively, calculated on the basis of $100$ stochastic numerical simulations. The solid curves represent the results obtained from the Euler method \cite{atkinson1989introduction, strogatz2018nonlinear} applied to Eqs. (\ref{eq1}) and (\ref{eq2}) with step size $0.01$. We terminate the Euler method when $t$ exceeds $300$ and the difference between the current time step and the last time step in terms of the fraction of infectious individuals is smaller than $10^{-9}$ for the first time.}\label{figsimulation}
\end{figure}

To assess the accuracy of Eqs. (\ref{eq1}) and (\ref{eq2}) to describe stochastic SIS dynamics, we run stochastic numerical simulations of the SIS model with the node2vec mobility on a network generated by the Barabási–Albert (BA) model \cite{barabasi1999emergence} with $100$ nodes. We assume that one individual that is selected uniformly at random is initially infectious and that all the other individuals are initially susceptible. The simulation results are shown by the circles and error bars in Fig. \ref{figsimulation}. The solid curves in Fig. \ref{figsimulation} indicate the fraction of infectious individuals in the equilibrium that we have obtained by integrating Eqs. (\ref{eq1}) and (\ref{eq2}). We confirm that the results obtained from stochastic numerical simulations are sufficiently close to those obtained from Eqs. (\ref{eq1}) and (\ref{eq2}). Therefore, we will exclusively use Eqs. (\ref{eq1}) and (\ref{eq2}) in the following analyses.

\subsection{Derivation of the epidemic threshold}

To derive the epidemic threshold, we rewrite Eqs. (\ref{eq1}) and (\ref{eq2}) as follows:
\begin{align}
    \frac{d\boldsymbol \rho_{\mathcal{S}}}{dt} = -\beta \boldsymbol \rho_{\mathcal{S}} \circ M \boldsymbol \rho_{\mathcal{I}} + \mu \boldsymbol \rho_{\mathcal{I}} - D_\mathcal{S} \boldsymbol \rho_{\mathcal{S}} + D_\mathcal{S} T^\top \boldsymbol \rho_{\mathcal{S}},
    \label{eq4}
\end{align}
\begin{align}
    \frac{d\boldsymbol \rho_{\mathcal{I}}}{dt}= \phantom{-}\beta \boldsymbol \rho_{\mathcal{S}} \circ M \boldsymbol \rho_{\mathcal{I}} - \mu \boldsymbol \rho_{\mathcal{I}} - D_\mathcal{I} \boldsymbol \rho_{\mathcal{I}} + D_\mathcal{I} T^\top \boldsymbol \rho_{\mathcal{I}},
    \label{eq5}
\end{align}
where $\boldsymbol \rho_{\mathcal{S}}=(\rho_{\mathcal{S},1}, \ldots, \rho_{\mathcal{S},2M})^\top$, $\boldsymbol \rho_{\mathcal{I}}=(\rho_{\mathcal{I},1}, \ldots, \rho_{\mathcal{I},2M})^\top$, $^{\top}$ represents the transposition, $\circ$ represents the Hadamard product, and $(M_{ij})$ is a $2M\times 2M$ matrix given by
\begin{align}
    M_{ij}=
    \begin{cases}
            1 & \mathrm{if}\ e_i(1)=e_j(1),\\
            0 & \mathrm{otherwise}.
    \end{cases}
\end{align}

Let $p_i^* (\mathrm{with\ } i=1, \ldots, 2M)$ be the stationary probability for directed edge $e_i$, i.e., 
\begin{align}
    (p^*_{1}, \ldots, p^*_{2M})T=(p^*_{1}, \ldots, p^*_{2M}).
\end{align}
In other words, $p_i^*$ is the stationary probability that an individual visits subpopulation $e_i(1)$ and the same individual visited subpopulation $e_i(0)$ just before $e_i(1)$. The disease-free equilibrium is given by 
\begin{align}
    (\boldsymbol \rho_{\mathcal{S}}^*; \boldsymbol \rho_{\mathcal{I}}^*)=\rho N(p_1^*, \ldots, p_{2M}^*; 0,\ldots, 0).
\end{align}
We apply a standard linearization technique to analyze the stability of the nonlinear dynamics represented by Eqs. (\ref{eq4}) and (\ref{eq5}) at the equilibrium $(\boldsymbol \rho_{\mathcal{S}}^*; \boldsymbol \rho_{\mathcal{I}}^*)$. The stability is determined by the eigenvalues of the Jacobian matrix \cite{walter1970ordinary, strogatz2018nonlinear}. We examine the Jacobian matrix of the nonlinear dynamics given by Eqs. (\ref{eq4}) and (\ref{eq5}) at the disease-free equilibrium. It is given by 
\begin{align}
    J=\begin{pmatrix}-D_\mathcal{S}L^\top & J_{12}\\ 0 & J_{22}\end{pmatrix},
\end{align}
where $L$ is the random-walk normalized Laplacian matrix \cite{masuda2017random}, i.e., 
\begin{align}
    L = I-T,
\end{align}
$I$ is the $2M \times 2M$ identity matrix,
\begin{align}
    J_{12}=-\beta \rho N \mathrm{diag}(p^*_{1}, \ldots, p^*_{2M}) M + \mu I,
    \label{eq10}
\end{align}
and
\begin{align}
    J_{22}=\beta \rho N \mathrm{diag}(p^*_{1}, \ldots, p^*_{2M}) M - \mu I - D_\mathcal{I} I+ D_\mathcal{I} T^\top.
    \label{eq14}
\end{align}
In Eqs. (\ref{eq10}) and (\ref{eq14}), $\mathrm{diag}()$ represents the diagonal matrix whose diagonal elements are given by the arguments. Matrix $J$ is isospectral to
\begin{align}
    \Tilde{J}=\begin{pmatrix}-D_\mathcal{S}L^\top & 0\\ 0 & J_{22}\end{pmatrix}.
\end{align}
Because $L$ is a Laplacian matrix, all the eigenvalues of $L$ have non-negative real parts. Because we have assumed that the original network of subpopulations is connected, the directed network induced by matrix $T$ is strongly connected, which implies that matrix $T$ is irreducible. By the Perron–Frobenius theorem, matrix $L$ has $0$ as a simple eigenvalue, i.e., the multiplicity of eigenvalue $0$ is equal to $1$. Because this zero eigenvalue corresponds to the probability flow of the node2vec random walk in the stationarity, it does not affect the stability of the disease-free equilibrium or the determination of the epidemic threshold. Therefore, the epidemic threshold $\beta_c$ is given by
\begin{align}
    \beta_c=\mathrm{max} \{\beta \ |\ \mathrm{max}(\Re(\mathrm{spec}(J_{22}))) = 0 \},
    \label{eq 14}
\end{align}
where $\mathrm{spec}(J_{22})$ denotes the spectrum (i.e., the set of all the eigenvalues) of matrix $J_{22}$, $\Re(\mathrm{spec}(J_{22}))$ is the set of real parts of the complex numbers belonging to $\mathrm{spec}(J_{22})$, and $\mathrm{max}$ denotes the maximum value.

\subsection{Epidemic threshold for various networks}

In this section, we numerically examine the epidemic threshold given by Eq. (\ref{eq 14}) as a function of the weight of backtracking, $a$, the weight of visiting the common neighbor of the currently visited and last visited subpopulations, $b$, and the diffusion rate for infectious individuals, $D_\mathcal{I}$, for the networks introduced in sections \ref{Synthetic networks} and \ref{Empirical networks}. Note that the epidemic threshold is independent of the diffusion rate for susceptible individuals, $D_\mathcal{S}$. Because simultaneously multiplying $\beta$, $\mu$, $D_\mathcal{S}$, and $D_\mathcal{I}$ by a positive constant $c$ is equivalent to replacing $t$ by $ct$ and not changing $\beta$, $\mu$, $D_\mathcal{S}$, or $D_\mathcal{I}$, we set $\mu=1$ without loss of generality. Equation (\ref{eq14}) indicates that there are three parameters $a, b$ and $D_\mathcal{I}$ that determine the epidemic threshold. Infection rate $\beta$ and average number of individuals per subpopulation $\rho$ only occur as their product. Therefore, we set $\rho=1$ without loss of generality. 

We use the bisection method to find the epidemic threshold given by Eq. (\ref{eq 14}). More precisely, given parameters $a, b,$ and $D_\mathcal{I}$, we set $\beta_\text{lower}=0.01$ and $\beta_\text{upper}=1.5$, which guarantee that $\mathrm{max}(\Re(\mathrm{spec}(J_{22})))<0$ and $\mathrm{max}(\Re(\mathrm{spec}(J_{22})))>0$, respectively, for all the following simulations. The eigenvalues are continuous functions of the entries of the matrix, which guarantees that $\mathrm{spec}(J_{22})$ is a continuous map in terms of $\beta$ \cite{ostrowski1973solutions, cucker1989alternate}. Because the composition of continuous maps is still continuous, the function $\mathrm{max}(\Re(\mathrm{spec}(J_{22})))$ is continuous in terms of $\beta$. Therefore, the bisection method converges to a root. With tolerance $\epsilon=10^{-4}$, we ran the bisection method with the initial bracketing interval $[\beta_\text{lower}, \beta_\text{upper}]$ to obtain the epidemic threshold $\beta_c$ with a truncation error less than $\epsilon$. 

In Fig. \ref{fig4}(a), we show the epidemic threshold for the Erdős–Rényi (ER) random graph with average degree $\langle k \rangle = 6$, $D_\mathcal{I}=1$ and various values of $a$ and $b$. We observe that the epidemic threshold increases as $a$ or $b$ decreases, which suggests that epidemic spreading is suppressed when the individuals travel without frequent backtracking or frequent visiting to common neighbors of the presently visited subpopulation and the last visited subpopulation. We observe that the dynamic range of the epidemic threshold $\beta_c$ when we fix $a$ and vary $b\in [0, 5]$ is larger than when we fix $b$ and vary $a\in [0, 5]$. Therefore, the weight of visiting the common neighbors more strongly influences the epidemic threshold than the weight of backtracking. The epidemic threshold when we set $b=1$ and vary $a$ and $D_\mathcal{I}$ and when we set and $a=1$ and vary $b$ and $D_\mathcal{I}$ is shown in Fig. \ref{fig4}(b) and Fig. \ref{fig4}(c), respectively. These two figures indicate that a larger diffusion rate $D_\mathcal{I}$ of the infectious individuals suppresses epidemic spreading, which is consistent with the previous results for the first-order mobility rule (i.e., simple random walk) on the network of subpopulations \cite{masuda2010effects, gomez2018critical, soriano2018spreading}. Figures \ref{fig4}(b) and \ref{fig4}(c) also indicate that changes in $D_\mathcal{I}\in [0, 10]$ have a larger impact on the epidemic threshold than changes in $a\in [0, 5]$ and $b\in [0, 5]$, whereas these parameters have different units.

\begin{figure}[!b]
  \includegraphics[width=1.02\textwidth]{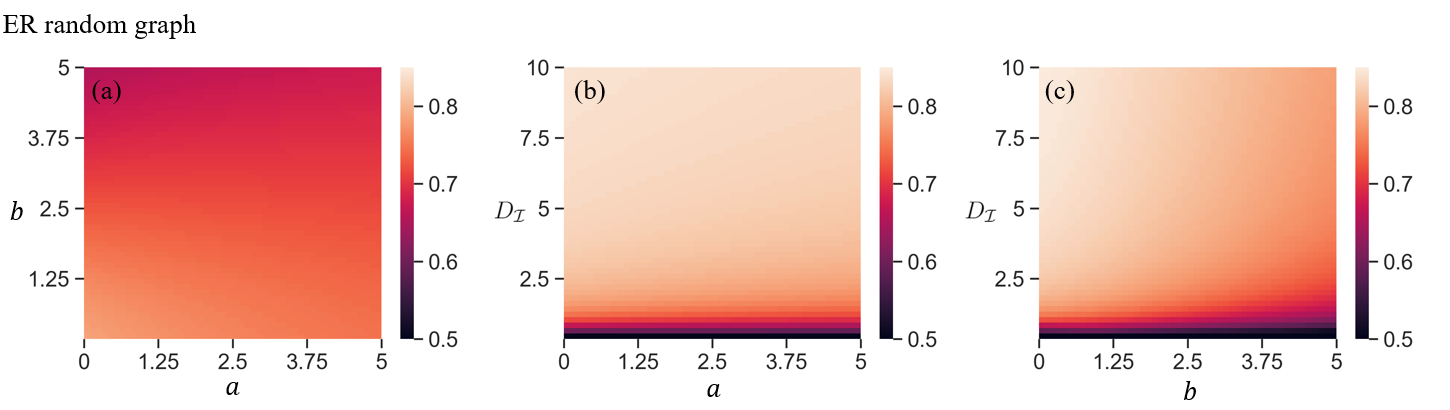}
  \includegraphics[width=1.02\textwidth]{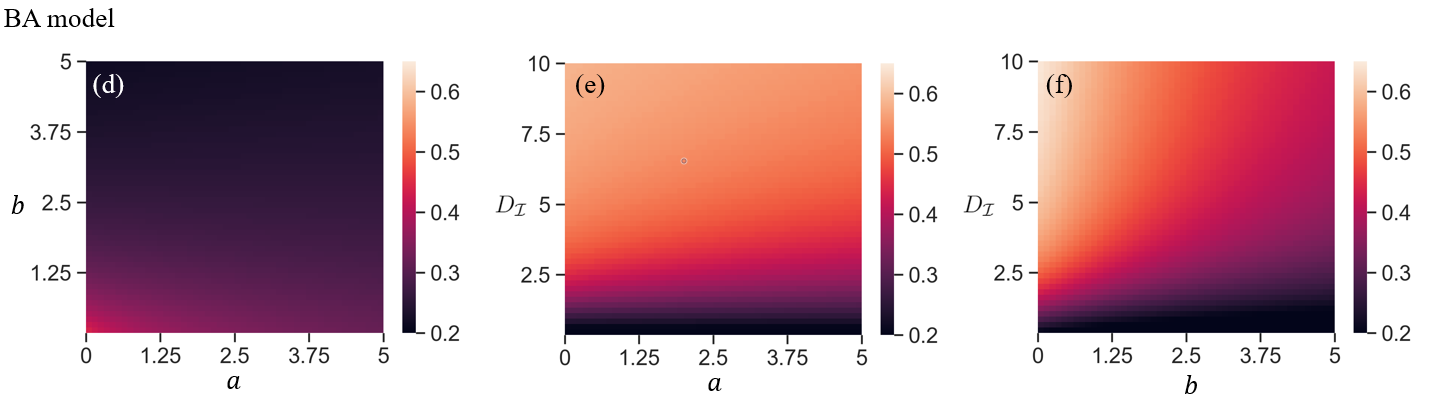}
  \includegraphics[width=1.02\textwidth]{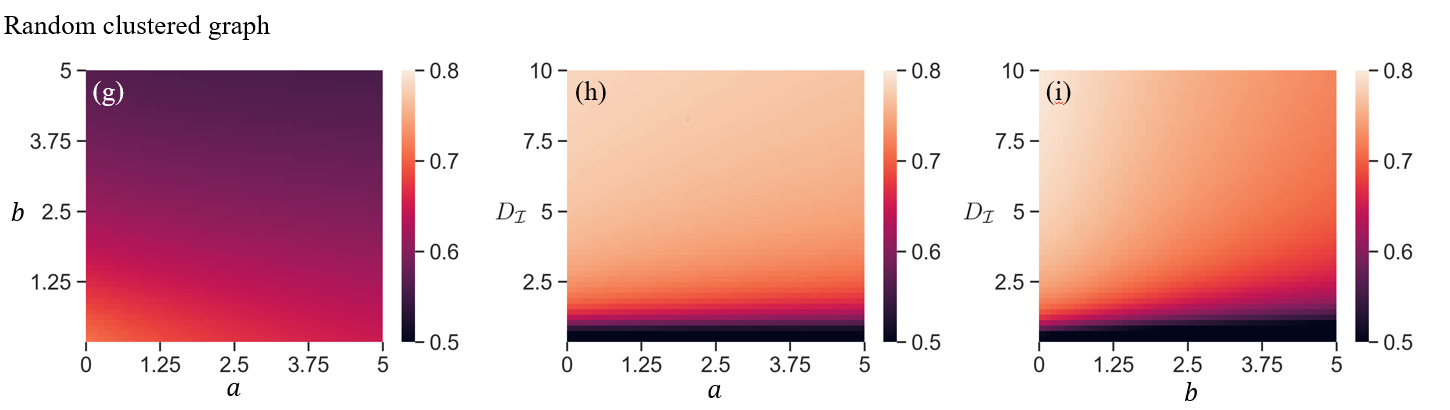}
  \caption{Epidemic threshold for the SIS model on the synthetic metapopulation networks. (a)--(c) ER random graph. (d)--(f) BA model. (g)--(i) Random clustered graph. (j)--(l) Power-law cluster graph. (m)--(o) Geographical threshold graph model. (p)--(r) LFR model. The hue represents the value of the epidemic threshold. Panels (a), (d), (g), (j), (m), and (p) represent the epidemic threshold $\beta_c$ as a function of $a$ and $b$ with $D_\mathcal{I} = 1$. Panels (b), (e), (h), (k), (n), and (q) represent the epidemic threshold $\beta_c$ as a function of $a$ and $D_\mathcal{I}$ with $b = 1$. Panels (c), (f), (i), (l), (o) and (r) represent the epidemic threshold $\beta_c$ as a function of $b$ and $D_\mathcal{I}$ with $a = 1$.}\label{fig4}
  \vspace*{-7pt}
\end{figure}

\begin{figure}[!t]
  \ContinuedFloat
  \includegraphics[width=1.02\textwidth]{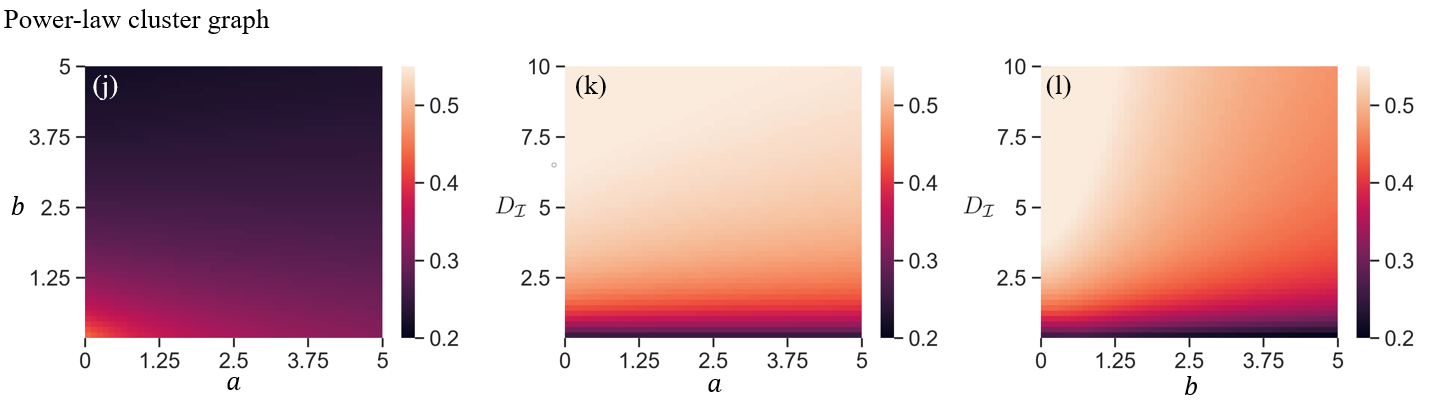}
  \includegraphics[width=1.02\textwidth]{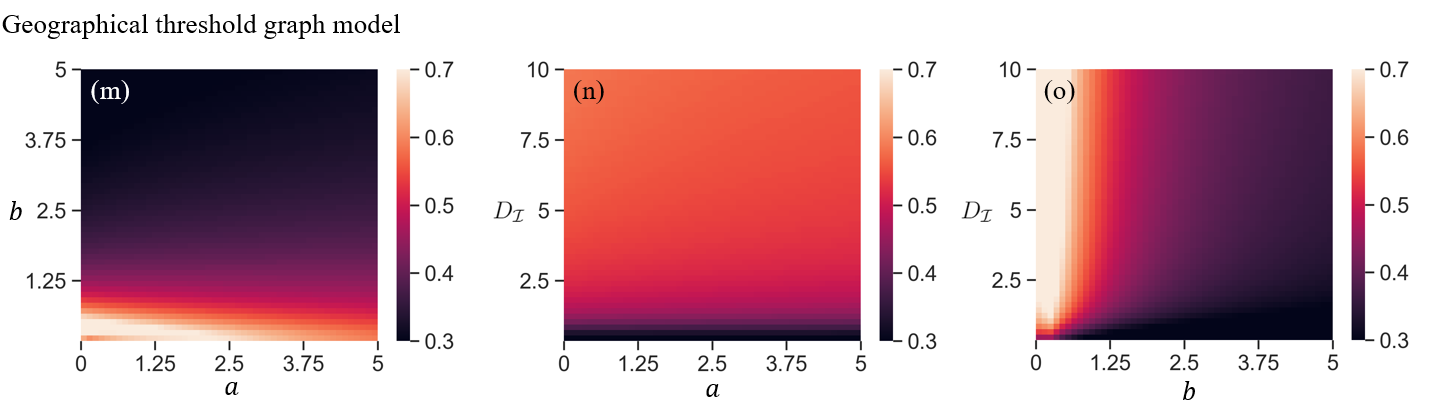}
  \includegraphics[width=1.02\textwidth]{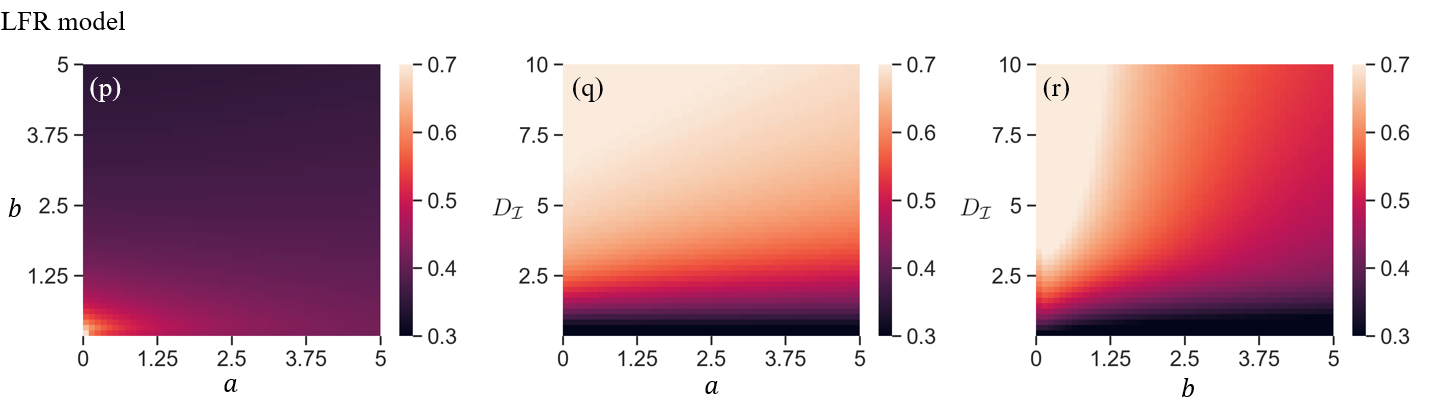}
  \caption*{Figure 4 (continued) }
  \vspace*{-7pt}
\end{figure}

We show the epidemic threshold for the BA model \cite{barabasi1999emergence} in Figs.~\ref{fig4}(d)--(f), random clustered graph \cite{miller2009percolation, newman2009random} in Figs.~\ref{fig4}(g)--(i), power-law cluster graph \cite{holme2002growing} in Figs.~\ref{fig4}(j)--(l), geographical threshold graph model \cite{masuda2005geographical} in Figs.~\ref{fig4}(m)--(o), and Lancichinetti–Fortunato–Radicchi (LFR) model \cite{lancichinetti2008benchmark} in Figs.~\ref{fig4}(p)--(r), for the same parameter regions as those used for the ER random graph. Figures \ref{fig4}(d), (g), (j), (m), and (p), in which we vary $a$ and $b$ with $D_\mathcal{I}=1$, suggest that the epidemic threshold largely decreases as $a$ or $b$ increases for all model networks, which is similar to the results for the ER random graph shown in Fig. \ref{fig4}(a). With an exception of the geographical threshold graph model at small $a$ and $b$ values (see Fig. \ref{fig4}(m)), the epidemic threshold is largest near $a=b=0$. We also find for these networks that the impact of $b$ on the epidemic threshold is stronger than that of $a$, which is consistent with Fig.~\ref{fig4}(a). Figures \ref{fig4}(b), (c), (e), (f), (h), (i), (k), (l), (n), (o), (q), and (r) indicate that a large diffusion rate suppresses epidemic spreading. Figures \ref{fig4}(b), (e), (h), (k), (n), and (q) indicate that the effect of $a$ on the epidemic threshold is much smaller than that of $D_\mathcal{I}$. These results are consistent with those for the ER random graph. In contrast, the effect of $b$ is smaller than $D_\mathcal{I}$ in some cases (Figs. \ref{fig4}(i) and (l)), which is similar to the case of the ER random graph (Fig. \ref{fig4}(c)), but is comparable to that of $D_\mathcal{I}$ for some networks (Fig. \ref{fig4}(f), (o), (r)). 

The observation above may fail for networks with a homogeneous degree distribution. To examine this possibility, we consider an extended ring network shown in Fig. \ref{figextend}, in which each node is connected to two neighbors on each side of the ring. For this network, we have derived the following theorem.

\bigskip

\textbf{Theorem 1}: The epidemic threshold for the extended ring network is equal to $\mu/\rho$, independently of the value of $a$, $b$, and $D_\mathcal{I}$.

\bigskip

\begin{figure}[!b]
  \includegraphics[width=1.02\textwidth]{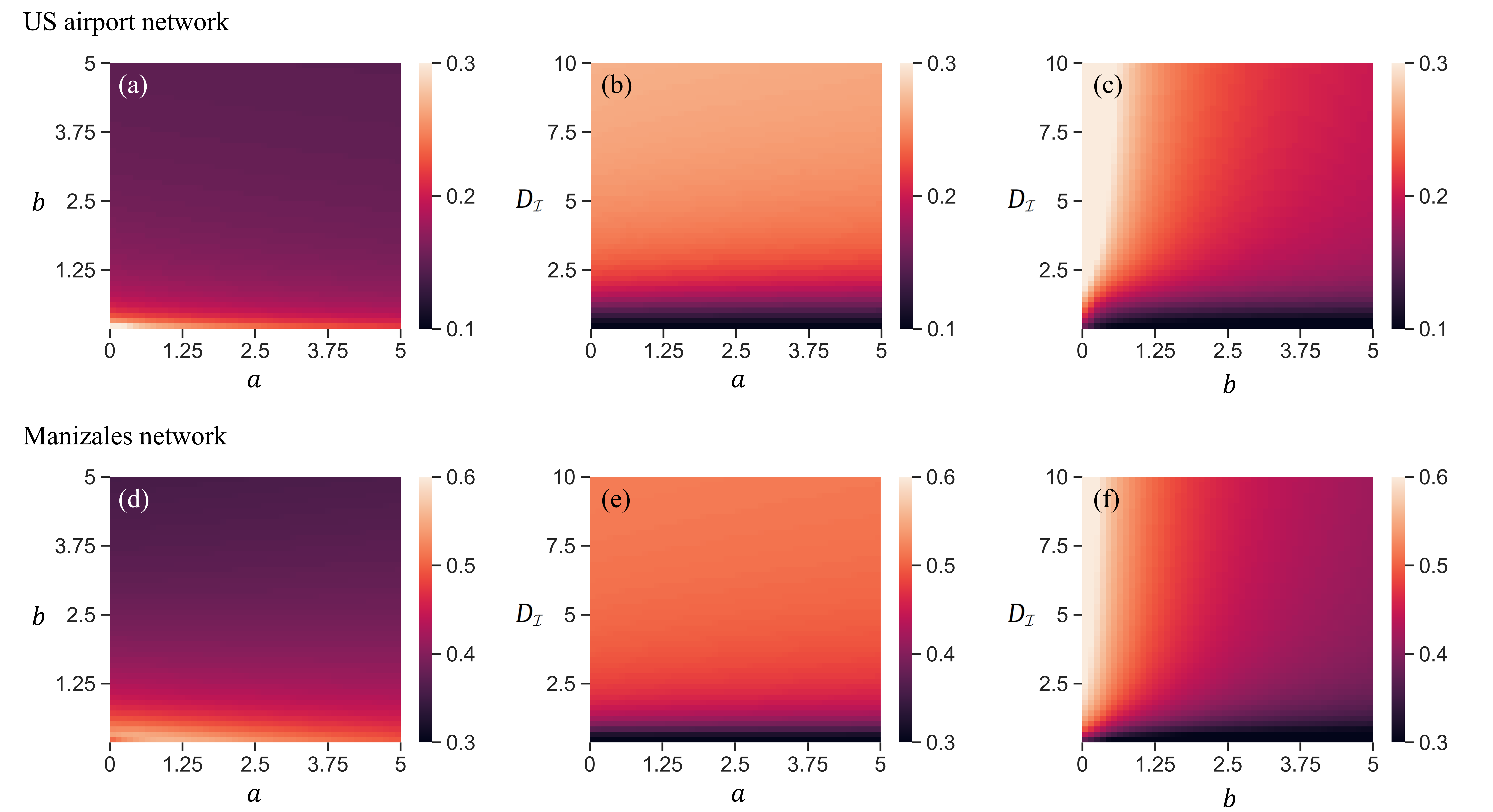}
  \caption{Epidemic threshold for the SIS model on the empirical metapopulation networks. (a)--(c) US airport network. (d)--(f) The wealthiest layer of the Manizales network. The hue represents the value of the epidemic threshold. Panels (a) and (d) represent the epidemic threshold $\beta_c$ as a function of $a$ and $b$ with $D_\mathcal{I}=1$. Panels (b) and (e) represent the epidemic threshold $\beta_c$ as a function of $a$ and $D_\mathcal{I}$ with $b = 1$. Panels (c) and (f) represent the epidemic threshold $\beta_c$ as a function of $b$ and $D_\mathcal{I}$ with $a = 1$.}\label{fig5}
  \vspace*{-7pt}
\end{figure}

Note that the statement of the Theorem does not hold true in general if the network is regular (i.e., a network in which all the nodes have the same degree) but is different from the extended ring network. The reason for this difference is probably that the extended ring network is not only regular but also vertex-transitive \cite{biggs1993algebraic, godsil2013algebraic, meng2020analysis}; casually speaking, all the nodes are equivalent to each other. We provide the proof of Theorem 1 in Appendix. This result contrasts to the numerical results shown in Figs.~\ref{fig4} and \ref{fig5}, in which the epidemic threshold systematically depends on the $a$, $b$, and $D_\mathcal{I}$ values. The reason for this difference is probably that the networks used in Figs.~\ref{fig4} and \ref{fig5} are relatively heterogeneous in terms of the node's degree, whereas all the nodes have the same degree in the extended ring network.

We find similar results for two empirical networks. The dependence of the epidemic threshold on $a$, $b$, and $D_\mathcal{I}$ for these two networks is shown in Fig. \ref{fig5}. The results are similar to those for the model networks shown in Fig. \ref{fig4}.

\section{Conclusions}

We investigated the SIS dynamics on metapopulation networks with the node2vec mobility. The node2vec is a second-order random walk, which one can transform into a first-order Markov chain by considering the set of all the directed edges and transitions among them \cite{rosvall2014memory, scholtes2014causality, meng2020analysis}. We built the master equations for the SIS dynamics involving the transition probability matrix of the node2vec random walk, $T$. We derived the epidemic threshold for any connected undirected networks. We observed that, for various synthetic and empirical networks with heterogeneous degree distributions, the epidemic threshold was large (and therefore, epidemic spreading is suppressed) when individuals explore the network without frequently backtracking (i.e., small $a$) or without frequently visiting common neighbors of the currently visited and the last visited subpopulations (i.e., small $b$).

The convergence of the node2vec random walk towards the stationary density is fast when $a$ or $b$ is small \cite{meng2020analysis}. In the present study, we have shown that the epidemic threshold tends to be large for small $a$ or $b$ values. Therefore, a fast convergence in the node2vec random walk is associated with a large epidemic threshold. When individuals obey the simple random walk in metapopulation models, the epidemic threshold for the SIS dynamics is known to increase as the diffusion rate increases \cite{masuda2010effects, gomez2018critical, soriano2018spreading}, which is consistent with our present numerical results. Because the rate of the convergence is generally proportional to the diffusion rate, a fast convergence in the simple random walk is associated with a large epidemic threshold. Therefore, our present result that the epidemic threshold tends to be large for small $a$, small $b$, or large $D_\mathcal{I}$ values collectively supports the idea that the epidemic threshold is large when the random walk rapidly converges towards the stationary density.

When individuals move according to Markovian random walks with recurrent mobility, the epidemic threshold does not necessarily increase monotonically as a function of the diffusion rate, and the relationship between the epidemic threshold and the diffusion rate depends on which subpopulations the individuals use as their home subpopulation \cite{gomez2018critical, soriano2018spreading, soriano2020impact}. We did not find such a nonmonotonicity because our models do not describe recurrent mobility with which different individuals use different subpopulations as home. It may be interesting to combine the recurrent mobility modeling and the node2vec random walk. Then, we may observe nonmonotonic dependence of the epidemic threshold on the $a$ or $b$ values as well as on the diffusion rate. This warrants future work.

Let us mention other possible directions for future research. First, we may be able to exploit our observation that the node2vec mobility with small $a$ and $b$ suppresses the spread of infections to inform intervention methods. A family of methods to intervene into epidemic dynamics on networks of subpopulations is to lessen the infection rate within subpopulations having large degrees \cite{tanaka2014random, matsuki2019intervention, gong2019modelling}. Under the simple random walk, containment of the epidemic has also been examined in combination with adaptive dynamics, with which individuals cancel their travel in response to their infection status \cite{meloni2011modeling, poletto2013human} or adjust their movement based on the information about the safety level measured by the number of susceptible individuals in various subpopulations \cite{meloni2011modeling, wang2012safety, nicolaides2013price, wang2017interplay}. Other containment methods include the shutdown of some edges connecting subpopulations \cite{hufnagel2004forecast, ferguson2006strategies}, usage of antibiotics and antiviral drugs \cite{ferguson2006strategies, colizza2007modeling}, and finding the most influential spreaders \cite{kitsak2010identification, ahajjam2018identification}. It is worth considering these containment methods assuming the node2vec mobility because the node2vec random walk may change the efficiency of these intervention methods. Furthermore, forcing individuals to use small $a$ or $b$ values itself may also be used for intervention, which may be combined with the aforementioned intervention strategies. Second, in some multilayer networks, the epidemic threshold is not a smooth function of the diffusion rate when the interlayer coupling is weak \cite{soriano2018spreading}. Examining the possibility of similar non-smooth behavior of the epidemic threshold in terms of the $a$ and $b$ under the node2vec mobility may be interesting. Finally, it is straightforward to extend our modeling framework to analyze the effects of the node2vec mobility on other dynamical processes on metapopulation model networks, such as the susceptible-infectious-recovered (SIR) model \cite{colizza2007invasion, matsuki2019intervention}, evolutionary games \cite{nagatani2018metapopulation, kabir2019evolutionary}, and prey-predator dynamics \cite{bonsall2004demographic, cooper2012intermediate}.

\section*{Acknowledgments}

N.M. acknowledges support from AFOSR European Office (under grant no. FA9550–19–1–7024), the Nakatani Foundation, the Sumitomo Foundation, and Japan Science and Technology Agency (JST) Moonshot R\&D (under grant no. JPMJMS2021).

\section*{Appendix: Proof of Theorem 1}

The outline of the following proof is to show that, as $\beta$ increases through $\beta = \mu / \rho$, there is an eigenvalue of $J_{22}$ whose real part changes from negative, to zero, and then to positive.

\begin{figure}[!h]
  \centering
  \includegraphics[width=0.55\textwidth]{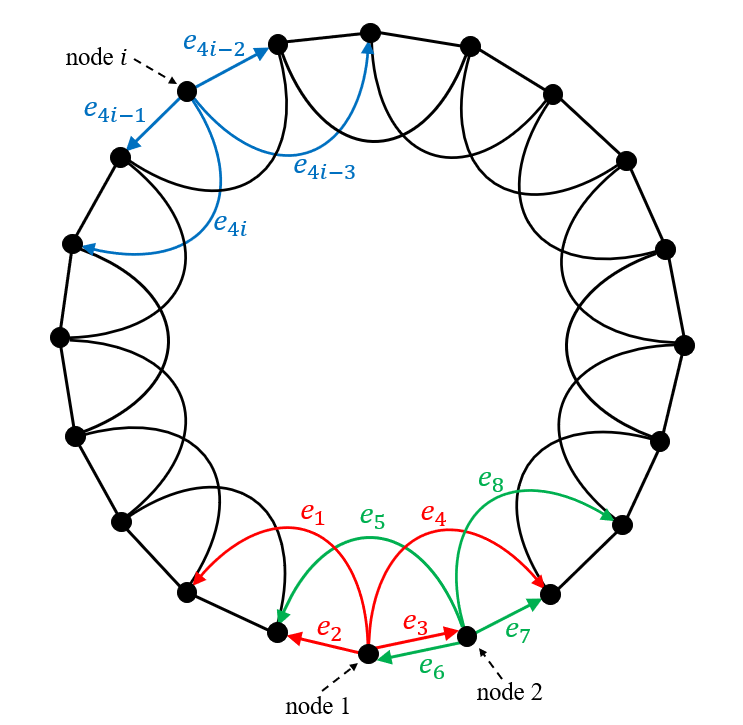}
  \caption{Labeling of the directed edges in the extended ring network. The nodes and the corresponding directed edges are labeled counterclockwise. }\label{figextendcolor}
  \vspace*{-7pt}
\end{figure}

We first present an explicit expression for matrix $J_{22}$ when the network is an extended ring. Given $k\times k$ matrices $B_i$, where $i=1,2,\ldots ,n$, we define the $kn\times kn$ block circulant matrix $\mathrm{bcirc}(B_1,B_2,\ldots, B_n)$ by
\begin{align}
    \mathrm{bcirc}(B_1,B_2,\ldots,B_n):=
    \begin{pmatrix}
    B_1 & B_2 & \cdots & B_{n-1} & B_n \\
    B_n & B_1 & B_2 & \cdots & B_{n-1} \\
    \vdots  & B_n & B_1 & \ddots & \vdots  \\
    B_3 & \ & \ddots & \ddots & B_2\\
    B_2 & B_3 & \cdots & B_n & B_1
    \end{pmatrix}.
\end{align}
We order the directed edges in $E$ in the extended ring network as illustrated in Fig. \ref{figextendcolor}. Then, the transition probability matrix, $T$, is block circulant and given by 

\begin{align}
    T=\mathrm{bcirc}(0,T_1,T_2,0,\ldots,0,T_3,T_4),
    \label{eqmatrixT}
\end{align}
where 
\begin{align}
    T_1=\frac{1}{a+2b+1}\begin{pmatrix}0&0&0&0 \\ 0&0&0&0\\ b&a&b&1\\ 0&0&0&0\end{pmatrix},
\end{align}
\begin{align}
    T_2=\frac{1}{a+b+2}\begin{pmatrix}0&0&0&0\\ 0&0&0&0\\ 0&0&0&0\\ a&b&1&1\end{pmatrix},
\end{align}
\begin{align}
    T_3=\frac{1}{a+b+2}\begin{pmatrix}1&1&b&a\\ 0&0&0&0\\ 0&0&0&0\\ 0&0&0&0\end{pmatrix},
\end{align}
and
\begin{align}
    T_4=\frac{1}{a+2b+1}\begin{pmatrix}0&0&0&0\\ 1&b&a&b\\ 0&0&0&0\\ 0&0&0&0\end{pmatrix}.
\end{align}
In this case, matrix
\begin{align}
    M=\mathrm{bcirc}(I_4, M_1, M_2, M_3, M_4, 0, \ldots, 0, M_4^\top, M_3^\top, M_2^\top, M_1^\top),
    \label{eqmatrixM}
\end{align}
is also block circulant, where 
\begin{align}
    I_4=\begin{pmatrix}1&0&0&0\\ 0&1&0&0\\ 0&0&1&0\\0&0&0&1\end{pmatrix},
\end{align}
\begin{align}
    M_1=\begin{pmatrix}0&0&0&0\\ 1&0&0&0\\ 0&0&0&0\\0&0&1&0\end{pmatrix},
\end{align}
\begin{align}
    M_2=\begin{pmatrix}0&0&0&0\\ 0&0&0&0\\ 0&1&0&0\\0&0&0&0\end{pmatrix},
\end{align}
\begin{align}
    M_3=\begin{pmatrix}0&0&0&0\\ 0&0&0&0\\ 1&0&0&0\\0&1&0&0\end{pmatrix},
\end{align}
and
\begin{align}
    M_4=\begin{pmatrix}0&0&0&0\\ 0&0&0&0\\ 0&0&0&0\\1&0&0&0\end{pmatrix}.
\end{align}
We also obtain
\begin{align}
    \mathrm{diag}(p^*_{1}, \ldots, p^*_{2M}) = \frac{1}{\mathcal{N}} \mathrm{bcirc}(D, 0, \ldots, 0), 
    \label{eqprobability}
\end{align}
where 
\begin{align}
    D=\begin{pmatrix}1&0&0&0\\ 0&\frac{a+2b+1}{a+b+2}&0&0\\ 0&0&\frac{a+2b+1}{a+b+2}&0\\0&0&0&1\end{pmatrix}
\end{align}
and $\mathcal{N}$ is the normalization constant. By combining Eqs. (\ref{eq14}), (\ref{eqmatrixT}), (\ref{eqmatrixM}), and (\ref{eqprobability}), we obtain
\begin{align}
\begin{split}
    J_{22} = & \frac{\beta\rho N}{\mathcal{N}}\mathrm{bcirc}(DI_4, DM_1, DM_2, DM_3, DM_4, 0, \ldots, 0, DM_4^\top, DM_3^\top, DM_2^\top, DM_1^\top) \nonumber \\ 
    & - (\mu + D_\mathcal{I})\mathrm{bcirc}(I_4, 0, \ldots, 0) + D_\mathcal{I}\mathrm{bcirc}(0,T_1,T_2,0,\ldots,0,T_3,T_4)^\top \nonumber \\ 
    = & \mathrm{bcirc} \left(\frac{\beta\rho N}{\mathcal{N}} DI_4 - (\mu + D_\mathcal{I})I_4, \frac{\beta\rho N}{\mathcal{N}}DM_1 + D_\mathcal{I}T_4^\top, \frac{\beta\rho N}{\mathcal{N}}DM_2 + D_\mathcal{I}T_3^\top, \frac{\beta\rho N}{\mathcal{N}}DM_3, \right.\\
    & \left. \frac{\beta\rho N}{\mathcal{N}}DM_4, 0, \ldots, 0, \frac{\beta\rho N}{\mathcal{N}}DM_4^\top, \frac{\beta\rho N}{\mathcal{N}}DM_3^\top, \frac{\beta\rho N}{\mathcal{N}}DM_2^\top + D_\mathcal{I}T_2^\top, \frac{\beta\rho N}{\mathcal{N}}DM_1^\top + D_\mathcal{I}T_1^\top \right )\\
    := & \mathrm{bcirc}(J_{22,0}, J_{22,1}, J_{22,2}, J_{22,3}, J_{22,4}, 0, \ldots, 0, J_{22,5}, J_{22,6}, J_{22,7}, J_{22,8}).
\end{split}
\end{align}

Next, let
\begin{align}
    \rho_j=e^{i\frac{2\pi j}{N}}
    \label{eq27}
\end{align}
denote the $N$th roots of $1$, where $i$ is the imaginary unit and $j=0,1,2,...,N-1$. Using $\rho_j$, we define $4\times 4$ matrices
\begin{align}
    H_j=J_{22,0} + J_{22,1}\rho_j + J_{22,2}\rho_j^2 + J_{22,3}\rho_j^3 + J_{22,4}\rho_j^4 + J_{22,5}\rho_j^{N-4} + J_{22,6}\rho_j^{N-3} + J_{22,7}\rho_j^{N-2} + J_{22,8}\rho_j^{N-1}.
    \label{eq28}
\end{align}
Theorem 3 in Ref. \cite{tee2007eigenvectors} guarantees that
\begin{align}
    \mathrm{spec}(J_{22})=\bigcup_{j=0}^{N-1} \mathrm{spec}(H_j).
    \label{eq23}
\end{align}
Therefore, the epidemic threshold $\beta_c$ is given by
\begin{align}
    \beta_c &=\mathrm{max} \{\beta \ |\ \mathrm{max}(\Re(\mathrm{spec}(J_{22})))=0\} \nonumber\\
    & = \mathrm{max} \{\beta \ |\ \mathrm{max}(\Re(\mathrm{spec}(H_j)))=0, \ j=0, \ldots , N-1 \}.
    \label{eq30}
\end{align}
The $\mathrm{max}$ in the second line of Eq. (\ref{eq30}) represents the maximum among the four eigenvalues of $H_j$ and among the $N$ values of $j$. The construction up to here is similar to that in our previous work for computing the spectral gap of node2vec random walks on networks \cite{meng2020analysis}.

To derive $\beta_c$, we further proceed as follows. Using Eq. (\ref{eq27}) and (\ref{eq28}), we obtain 
\begin{align}
    H_j = & \frac{\beta \rho}{s}\begin{pmatrix}
    1&\rho_j^{N-1}&\rho_j^{N-3}&\rho_j^{N-4}\\ 
    \frac{a+2b+1}{a+b+2} \rho_j &\frac{a+2b+1}{a+b+2}&\frac{a+2b+1}{a+b+2} \rho_j^{N-2}&\frac{a+2b+1}{a+b+2}\rho_j^{N-3}\\ 
    \frac{a+2b+1}{a+b+2}\rho_j^3 &\frac{a+2b+1}{a+b+2} \rho_j^2&\frac{a+2b+1}{a+b+2}&\frac{a+2b+1}{a+b+2}\rho_j^{N-1}\\ 
    \rho_j^4&\rho_j^3&\rho_j&1
    \end{pmatrix} \nonumber \\
    & + D_\mathcal{I}\begin{pmatrix}\frac{1}{a+b+2}\rho_j^2&\frac{1\:}{a+2b+1}\rho_j&\frac{\:b}{a+2b+1}\rho_j^{N-1}&\frac{\:a}{a+b+2}\rho_j^{N-2}\\ \frac{\:1}{a+b+2}\rho_j^2&\frac{b\:}{a+2b+1}\rho_j&\frac{\:a}{a+2b+1}\rho_j^{N-1}&\frac{\:b}{a+b+2}\rho_j^{N-2}\\ \frac{\:b}{a+b+2}\rho_j^2&\frac{\:a}{a+2b+1}\rho_j&\frac{\:b}{a+2b+1}\rho_j^{N-1}&\frac{\:1}{a+b+2}\rho_j^{N-2}\\ \frac{\:a}{a+b+2}\rho_j^2&\frac{\:b}{a+2b+1}\rho_j&\frac{\:1}{a+2b+1}\rho_j^{N-1}&\frac{\:1}{a+b+2}\rho_j^{N-2}\end{pmatrix} - (\mu+D_\mathcal{I})I_4,
\end{align}
where $j=0,\ldots, N-1$, and
\begin{align}
    s=2+\frac{2(a+2b+1)}{a+b+2}.
    \label{eq35}
\end{align}
In particular, we obtain
\begin{align}
    H_0=\frac{\beta \rho}{s}\begin{pmatrix}1&1&1&1\\ \frac{a+2b+1}{a+b+2}&\frac{a+2b+1}{a+b+2}&\frac{a+2b+1}{a+b+2}&\frac{a+2b+1}{a+b+2}\\ \frac{a+2b+1}{a+b+2}&\frac{a+2b+1}{a+b+2}&\frac{a+2b+1}{a+b+2}&\frac{a+2b+1}{a+b+2}\\ 1&1&1&1\end{pmatrix} + D_\mathcal{I}\begin{pmatrix}\frac{1}{a+b+2}&\frac{1\:}{a+2b+1}&\frac{\:b}{a+2b+1}&\frac{\:a}{a+b+2}\\ \frac{\:1}{a+b+2}&\frac{b\:}{a+2b+1}&\frac{\:a}{a+2b+1}&\frac{\:b}{a+b+2}\\ \frac{\:b}{a+b+2}&\frac{\:a}{a+2b+1}&\frac{\:b}{a+2b+1}&\frac{\:1}{a+b+2}\\ \frac{\:a}{a+b+2}&\frac{\:b}{a+2b+1}&\frac{\:1}{a+2b+1}&\frac{\:1}{a+b+2}\end{pmatrix} - (\mu +D_\mathcal{I})I_4.
\end{align}

The main objective of the remainder of the proof is to show that the eigenvalue of $J_{22}$ with the largest real part is completely determined by the matrix $H_0$. The sum of each column of $H_0$ is equal to $\beta\rho - \mu$. Therefore, we distinguish three cases and claim the following statements:

(S1) If $\beta < \mu/\rho$, then each Gershgorin circle of $H_0$ created by each of its columns is contained in the left half of the complex plane excluding the imaginary axis. Therefore, the real part of each eigenvalue of $H_0$ is negative, i.e., $\Re(\mathrm{spec}(H_0)) < 0$.

(S2) If $\beta = \mu/\rho$, then the row vector $\boldsymbol 1=(1, 1, 1, 1)$ is a left eigenvector of $H_0$ associated with eigenvalue $0$, i.e., $\boldsymbol 1 H_0 = \boldsymbol 0$.

(S3) If $\beta > \mu/\rho$, then $\beta\rho - \mu$ is a positive eigenvalue of $H_0$ with the associated eigenvector $\boldsymbol 1$.

To verify statement (S2), we evaluate each coordinate of row vector $\boldsymbol 1 H_0$:
\begin{align*}
    (\boldsymbol 1 H_0)_i = & \frac{\beta \rho}{s} \left(1 + \frac{a+2b+1}{a+b+2} + \frac{a+2b+1}{a+b+2} + 1\right) \\ 
    & + D_\mathcal{I} \left( \frac{1}{a+b+2} + \frac{1}{a+b+2} + \frac{b}{a+b+2} + \frac{a}{a+b+2}\right) - (\mu + D_\mathcal{I})\nonumber\\
    = & \frac{\beta \rho}{s} \left[2+\frac{2(a+2b+1)}{a+b+2} \right] + D_\mathcal{I} - (\mu + D_\mathcal{I})\nonumber\\
    = & \beta\rho + D_\mathcal{I} - (\mu + D_\mathcal{I})\nonumber\\
    = & 0,
\end{align*}
where $i=1, 2, 3, 4$. This finishes the proof of statement (S2). One can verify statement (S3) in the same manner.

To prove (S1), we assume that $\beta < \mu/\rho$. We consider the Gershgorin circle created by the first column of $H_0$ as an example. The coordinate of the center of this Gershgorin circle is $(C_1, 0)$, where
\begin{align}
    C_1 & = \frac{\beta \rho}{s}+\frac{D_\mathcal{I}}{a+b+2}-\mu-D_\mathcal{I} \nonumber \\
    & < \left(\frac{\mu}{s} - \mu \right) + \left(\frac{D_\mathcal{I}}{a+b+2}-D_\mathcal{I} \right) \nonumber \\
    & < 0. 
    \label{eq 371}
\end{align}
In the first inequality in Eq. (\ref{eq 371}), we used $\beta < \mu / \rho$. In the second inequality, we used $a>0, b>0$, and $s>2$; note that Eq. (\ref{eq35}) implies $s>2$. The radius of the corresponding Greshgorin circle, denoted by $R_1$, is given by
\begin{align}
    R_1 & = \frac{\beta \rho}{s} \left[1+\frac{2(a+2b+1)}{a+b+2}\right] + \frac{a+b+1}{a+b+2} D_\mathcal{I}.
\end{align}
Note that $C_1 + R_1$ is the sum of the first column of $H_0$, which is equal to $\beta \rho - \mu <0$. Therefore, the Gershgorin circle is contained in the left half of the complex plane without touching the imaginary axis. Similarity, it is straightforward to show that all the Gershgorin circles of $H_0$ created by its columns are contained in the left half of the complex plane without touching the imaginary axis. This completes the proof of (S1). 

\begin{figure}[!h]
  \centering
  \includegraphics[width=0.55\textwidth]{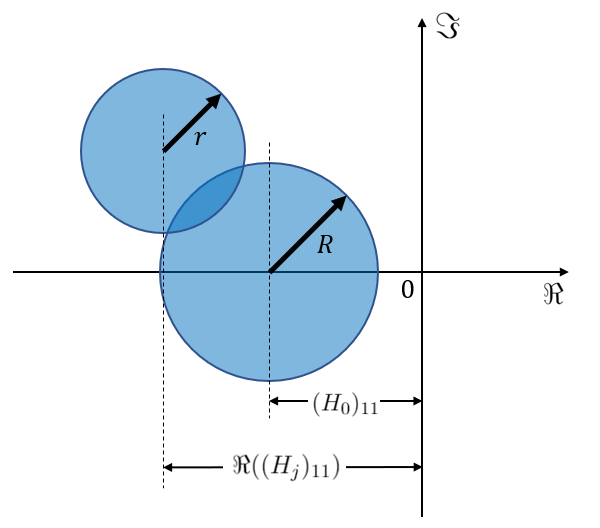}
  \caption{Geometric interpretation for Eqs. (\ref{eq37}) and (\ref{eq38}) when $\beta<\mu/\rho$. The larger Gershgorin circle with radius $R=\sum_{\ell=1; \ell \neq i}^4 |(H_0)_{\ell 1}|$ is centered at $((H_0)_{11}, 0)$. The smaller Gershgorin circle with radius $r=\sum_{\ell=1; \ell \neq i}^4 |(H_j)_{\ell 1}|$ is centered at $( \Re ((H_j)_{11}), \Im (\frac{D_\mathcal{I}}{a+b+2}\rho_j^2))$, where $\Re$ and $\Im$ represent the real and imaginary part of the complex number, respectively. }
  \label{fig6}
\end{figure}

Now we assess the Gershgorin circles generated by $H_j$ with $j=1, \ldots, N-1$. When $\beta < \mu/\rho$, the diagonal elements of $H_j$ satisfy 
\begin{align}
    \Re ((H_j)_{ii})<(H_0)_{ii}<0,
    \label{eq 37}
\end{align}
for any $i=1, \ldots, 4$ and $j=1, \ldots, N-1$, because multiplication by $\rho_j$ is a rotation about the origin in the complex plane. For example, one obtains
\begin{align}
    \Re ((H_j)_{11}) & = \frac{\beta \rho}{s} + \Re \left(\frac{D_\mathcal{I}}{a+b+2}\rho_j^2 \right) - (\mu +D_\mathcal{I}) \nonumber\\
    & < \frac{\beta\rho}{s}+\frac{D_\mathcal{I}}{a+b+2}-(\mu + D_\mathcal{I}) \nonumber \\
    & = (H_0)_{11}.
    \label{eq37}
\end{align}
Equation (\ref{eq 37}) implies that the center of the Gershgorin circle for $H_j$ created by its $i$th column is located to the left of the center of the corresponding Gershgorin circle for $H_0$ in the complex plane. It also holds true that 
\begin{align}
    \displaystyle \sum_{\ell =1; \ell \neq i}^4 |(H_j)_{\ell i}| < \sum_{\ell =1; \ell \neq i}^4 |(H_0)_{\ell i}|,
    \label{eq 38}
\end{align}
for any $i=1, \ldots, 4$ and $j=1, \ldots, N-1$. For example, one obtains
\begin{align}
    \sum_{\ell =1; \ell \neq i}^4 |(H_j)_{\ell 1}| & = \left|\frac{\beta\rho}{s} \frac{a+2b+1}{a+b+2} \rho_j + \frac{D_\mathcal{I}}{a+b+2}\rho_j^2 \right| + \left|\frac{\beta\rho}{s} \frac{a+2b+1}{a+b+2} \rho_j^3 + \frac{b D_\mathcal{I}}{a+b+2}\rho_j^2 \right| + \left|\frac{\beta\rho}{s} \rho_j^4 + \frac{a D_\mathcal{I}}{a+b+2}\rho_j^2 \right| \nonumber\\
    & < \left|\frac{\beta\rho}{s} \frac{a+2b+1}{a+b+2} \rho_j \right |+ \left |\frac{D_\mathcal{I}}{a+b+2}\rho_j^2 \right| + \left|\frac{\beta\rho}{s} \frac{a+2b+1}{a+b+2} \rho_j^3 \right |+ \left | \frac{b D_\mathcal{I}}{a+b+2}\rho_j^2 \right| + \left|\frac{\beta\rho}{s} \rho_j^4 \right |+ \left | \frac{a D_\mathcal{I}}{a+b+2}\rho_j^2 \right| \nonumber \\
    & = \left|\frac{\beta\rho}{s} \frac{a+2b+1}{a+b+2} \right |+ \left |\frac{D_\mathcal{I}}{a+b+2} \right| + \left|\frac{\beta\rho}{s} \frac{a+2b+1}{a+b+2}  \right |+ \left | \frac{b D_\mathcal{I}}{a+b+2} \right| + \left|\frac{\beta\rho}{s} \right |+ \left | \frac{a D_\mathcal{I}}{a+b+2} \right| \nonumber \\
    & = \sum_{\ell =1; \ell \neq i}^4 |(H_0)_{\ell 1}|.
    \label{eq38}
\end{align}
Equation (\ref{eq 38}) implies that the radius of the Gershgorin circle for $H_j$ created by its $i$th column is smaller than the radius of the corresponding Gershgorin circle for $H_0$. Because each Gershgorin circle for $H_j$, where $j=1, \ldots, N-1$, has its center to the left of that of the corresponding circle for $H_0$ and its radius is smaller than that of the corresponding circle for $H_0$, all the Gershgorin circles for $H_j$ (with $j=0, 1, \ldots, N-1$) are contained in the left half of the complex plane without touching the imaginary axis. Recall that $\mathrm{spec}(J_{22})=\bigcup_{j=0}^{N-1} \mathrm{spec}(H_j)$. Therefore, all the eigenvalues of $J_{22}$ have negative real part when $\beta<\mu/\rho$. This situation is schematically shown in Fig. \ref{fig6}. 

The case $\beta=\mu/\rho$ is similar to the case $\beta<\mu/\rho$. The only difference is that the four Gershgorin circles generated by $H_0$ are tangent to the imaginary axis. Therefore, we obtain $\mathrm{max}(\Re(\mathrm{spec}(H_j))) < 0$ for $j=1, \ldots, N-1$, and $\mathrm{max}(\Re(\mathrm{spec}(H_0))) \leq 0$. Furthermore, by statement (S2), $H_0$ has an eigenvalue $0$. Therefore, when $\beta=\mu/\rho$, all the eigenvalues of $J_{22}$ have negative real part except the zero eigenvalue. Statement (S3) implies that $J_{22}$ has a positive eigenvalue $\beta\rho-\mu$ when $\beta>\mu/\rho$. These results verify that $\beta_c=\mu/\rho$ is the epidemic threshold. 

\section*{Data accessibility}
The empirical network data sets are open resources and available at \cite{pajek2006, lotero2016rich}. The Python codes and synthetic network data sets used in the present study are available on Github \cite{code}.

\bibliographystyle{elsarticle-num}
\bibliography{Ref.bib}

\begin{thebibliography}{10}
\expandafter\ifx\csname url\endcsname\relax
  \def\url#1{\texttt{#1}}\fi
\expandafter\ifx\csname urlprefix\endcsname\relax\def\urlprefix{URL }\fi
\expandafter\ifx\csname href\endcsname\relax
  \def\href#1#2{#2} \def\path#1{#1}\fi

\bibitem{keeling2005networks}
M.~J. Keeling, K.~T. Eames, Networks and epidemic models, J. R. Soc. Interface
  2 (2005) 295--307.

\bibitem{danon2011networks}
L.~Danon, A.~P. Ford, T.~House, C.~P. Jewell, M.~J. Keeling, G.~O. Roberts,
  J.~V. Ross, M.~C. Vernon, Networks and the epidemiology of infectious
  disease, Interdiscip. Perspect. Infect. Dis. 2011 (2011) 284909.

\bibitem{masuda2013predicting}
N.~Masuda, P.~Holme, Predicting and controlling infectious disease epidemics
  using temporal networks, F1000Prime Rep. 5 (2013).

\bibitem{pellis2015eight}
L.~Pellis, F.~Ball, S.~Bansal, K.~Eames, T.~House, V.~Isham, P.~Trapman, Eight
  challenges for network epidemic models, Epidemics 10 (2015) 58--62.

\bibitem{pastor2015epidemic}
R.~Pastor-Satorras, C.~Castellano, P.~Van~Mieghem, A.~Vespignani, Epidemic
  processes in complex networks, Rev. Mod. Phys. 87 (2015) 925--979.

\bibitem{kiss2017mathematics}
I.~Z. Kiss, J.~C. Miller, P.~L. Simon, Mathematics of Epidemics on Networks,
  Springer, Cham, 2017.

\bibitem{hethcote1978immunization}
H.~W. Hethcote, An immunization model for a heterogeneous population, Theor.
  Popul. Biol. 14 (1978) 338--349.

\bibitem{may1984spatial}
R.~M. May, R.~M. Anderson, Spatial heterogeneity and the design of immunization
  programs, Math. Biosci. 72 (1984) 83--111.

\bibitem{lloyd1996spatial}
A.~L. Lloyd, R.~M. May, Spatial heterogeneity in epidemic models, J. Theor.
  Biol. 179 (1996) 1--11.

\bibitem{grenfell1997meta}
B.~Grenfell, J.~Harwood, ({M}eta) population dynamics of infectious diseases,
  Trends Ecol. Evol. 12 (1997) 395--399.

\bibitem{grenfell1998cities}
B.~Grenfell, B.~Bolker, Cities and villages: infection hierarchies in a measles
  metapopulation, Ecol. Lett. 1 (1998) 63--70.

\bibitem{hufnagel2004forecast}
L.~Hufnagel, D.~Brockmann, T.~Geisel, Forecast and control of epidemics in a
  globalized world, Proc. Natl. Acad. Sci. USA 101 (2004) 15124--15129.

\bibitem{colizza2006role}
V.~Colizza, A.~Barrat, M.~Barth{\'e}lemy, A.~Vespignani, The role of the
  airline transportation network in the prediction and predictability of global
  epidemics, Proc. Natl. Acad. Sci. USA 103 (2006) 2015--2020.

\bibitem{colizza2007reaction}
V.~Colizza, R.~Pastor-Satorras, A.~Vespignani, Reaction--diffusion processes
  and metapopulation models in heterogeneous networks, Nat. Phys. 3 (2007)
  276--282.

\bibitem{colizza2008epidemic}
V.~Colizza, A.~Vespignani, Epidemic modeling in metapopulation systems with
  heterogeneous coupling pattern: Theory and simulations, J. Theor. Biol. 251
  (2008) 450--467.

\bibitem{balcan2009multiscale}
D.~Balcan, V.~Colizza, B.~Gon{\c{c}}alves, H.~Hu, J.~J. Ramasco, A.~Vespignani,
  Multiscale mobility networks and the spatial spreading of infectious
  diseases, Proc. Natl. Acad. Sci. USA 106 (2009) 21484--21489.

\bibitem{masuda2010effects}
N.~Masuda, Effects of diffusion rates on epidemic spreads in metapopulation
  networks, New J. Phys. 12 (2010) 093009.

\bibitem{balcan2011phase}
D.~Balcan, A.~Vespignani, Phase transitions in contagion processes mediated by
  recurrent mobility patterns, Nat. Phys. 7 (2011) 581--586.

\bibitem{vespignani2012modelling}
A.~Vespignani, Modelling dynamical processes in complex socio-technical
  systems, Nat. Phys. 8 (2012) 32--39.

\bibitem{nicolaides2012metric}
C.~Nicolaides, L.~Cueto-Felgueroso, M.~C. Gonz{\'a}lez, R.~Juanes, A metric of
  influential spreading during contagion dynamics through the air
  transportation network, PLoS ONE 7 (2012) e40961.

\bibitem{tizzoni2015scaling}
M.~Tizzoni, K.~Sun, D.~Benusiglio, M.~Karsai, N.~Perra, The scaling of human
  contacts and epidemic processes in metapopulation networks, Sci. Rep. 5
  (2015) 15111.

\bibitem{gomez2018critical}
J.~G{\'o}mez-Garde{\~n}es, D.~Soriano-Pa{\~n}os, A.~Arenas, Critical regimes
  driven by recurrent mobility patterns of reaction--diffusion processes in
  networks, Nat. Phys. 14 (2018) 391--395.

\bibitem{soriano2018spreading}
D.~Soriano-Pa{\~n}os, L.~Lotero, A.~Arenas, J.~G{\'o}mez-Garde{\~n}es,
  Spreading processes in multiplex metapopulations containing different
  mobility networks, Phys. Rev. X 8 (2018) 031039.

\bibitem{soriano2020vector}
D.~Soriano-Pa{\~n}os, J.~H. Arias-Castro, A.~Reyna-Lara, H.~J. Mart{\'\i}nez,
  S.~Meloni, J.~G{\'o}mez-Garde{\~n}es, Vector-borne epidemics driven by human
  mobility, Phys. Rev. Research 2 (2020) 013312.

\bibitem{colizza2007invasion}
V.~Colizza, A.~Vespignani, Invasion threshold in heterogeneous metapopulation
  networks, Phys. Rev. Lett. 99 (2007) 148701.

\bibitem{belik2011natural}
V.~Belik, T.~Geisel, D.~Brockmann, Natural human mobility patterns and spatial
  spread of infectious diseases, Phys. Rev. X 1 (2011) 011001.

\bibitem{balcan2012invasion}
D.~Balcan, A.~Vespignani, Invasion threshold in structured populations with
  recurrent mobility patterns, J. Theor. Biol. 293 (2012) 87--100.

\bibitem{poletto2013human}
C.~Poletto, M.~Tizzoni, V.~Colizza, Human mobility and time spent at
  destination: impact on spatial epidemic spreading, J. Theor. Biol. 338 (2013)
  41--58.

\bibitem{rosvall2014memory}
M.~Rosvall, A.~V. Esquivel, A.~Lancichinetti, J.~D. West, R.~Lambiotte, Memory
  in network flows and its effects on spreading dynamics and community
  detection, Nat. Comm. 5 (2014) 4630.

\bibitem{scholtes2014causality}
I.~Scholtes, N.~Wider, R.~Pfitzner, A.~Garas, C.~J. Tessone, F.~Schweitzer,
  Causality-driven slow-down and speed-up of diffusion in non-markovian
  temporal networks, Nat. Comm. 5 (2014) 5024.

\bibitem{matamalas2016assessing}
J.~T. Matamalas, M.~De~Domenico, A.~Arenas, Assessing reliable human mobility
  patterns from higher order memory in mobile communications, J. R. Soc.
  Interface 13 (2016) 20160203.

\bibitem{granell2018epidemic}
C.~Granell, P.~J. Mucha, Epidemic spreading in localized environments with
  recurrent mobility patterns, Phys. Rev. E 97 (2018) 052302.

\bibitem{soriano2020impact}
D.~Soriano-Pa{\~n}os, G.~Ghoshal, A.~Arenas, J.~G{\'o}mez-Garde{\~n}es, Impact
  of temporal scales and recurrent mobility patterns on the unfolding of
  epidemics, J. Stat. Mech. 2020 (2020) 024006.

\bibitem{feng2020epidemic}
L.~Feng, Q.~Zhao, C.~Zhou, Epidemic spreading in heterogeneous networks with
  recurrent mobility patterns, Phys. Rev. E 102 (2020) 022306.

\bibitem{xuan2013reaction}
Q.~Xuan, F.~Du, L.~Yu, G.~Chen, Reaction-diffusion processes and metapopulation
  models on duplex networks, Phys. Rev. E 87 (2013) 032809.

\bibitem{wang2014epidemic}
B.~Wang, G.~Tanaka, H.~Suzuki, K.~Aihara, Epidemic spread on interconnected
  metapopulation networks, Phys. Rev. E 90 (2014) 032806.

\bibitem{grover2016node2vec}
A.~Grover, J.~Leskovec, node2vec: Scalable feature learning for networks, in:
  Proc. 22nd ACM SIGKDD International Conference on Knowledge Discovery and
  Data Mining, 2016, pp. 855--864.

\bibitem{qiu2018network}
J.~Qiu, Y.~Dong, H.~Ma, J.~Li, K.~Wang, J.~Tang, Network embedding as matrix
  factorization: Unifying deepwalk, line, pte, and node2vec, in: Proc. Eleventh
  ACM International Conference on Web Search and Data Mining, 2018, pp.
  459--467.

\bibitem{meng2020analysis}
L.~Meng, N.~Masuda, Analysis of node2vec random walks on networks, Proc. R.
  Soc. A 476 (2020) 20200447.

\bibitem{barabasi1999emergence}
A.-L. Barab{\'a}si, R.~Albert, Emergence of scaling in random networks, Science
  286 (1999) 509--512.

\bibitem{miller2009percolation}
J.~C. Miller, Percolation and epidemics in random clustered networks, Phys.
  Rev. E 80 (2009) 020901.

\bibitem{newman2009random}
M.~E. Newman, Random graphs with clustering, Phys. Rev. Lett. 103 (2009)
  058701.

\bibitem{holme2002growing}
P.~Holme, B.~J. Kim, Growing scale-free networks with tunable clustering, Phys.
  Rev. E 65 (2002) 026107.

\bibitem{masuda2005geographical}
N.~Masuda, H.~Miwa, N.~Konno, Geographical threshold graphs with small-world
  and scale-free properties, Phys. Rev. E 71 (2005) 036108.

\bibitem{lancichinetti2008benchmark}
A.~Lancichinetti, S.~Fortunato, F.~Radicchi, Benchmark graphs for testing
  community detection algorithms, Phys. Rev. E 78 (2008) 046110.

\bibitem{pajek2006}
Batagelj V and Mrvar A 2006 Pajek datasets
  \url{http://vlado.fmf.uni-lj.si/pub/networks/data/}. Accessed on Jan 6, 2021.

\bibitem{lotero2016rich}
L.~Lotero, R.~G. Hurtado, L.~M. Flor{\'\i}a, J.~G{\'o}mez-Garde{\~n}es, Rich do
  not rise early: Spatio-temporal patterns in the mobility networks of
  different socio-economic classes, Roy. Soc. Open Sci. 3 (2016) 150654.

\bibitem{atkinson1989introduction}
K.~E. Atkinson, An Introduction to Numerical Analysis, John wiley \& sons,
  United States, 1989.

\bibitem{strogatz2018nonlinear}
S.~H. Strogatz, Nonlinear Dynamics and Chaos with Student Solutions Manual:
  With Applications to Physics, Biology, Chemistry, and Engineering, 2nd
  Edition, CRC press, 2018.

\bibitem{walter1970ordinary}
W.~Walter, Ordinary differential equations, in: Differential and Integral
  Inequalities, Springer, 1970, pp. 63--123.

\bibitem{masuda2017random}
N.~Masuda, M.~A. Porter, R.~Lambiotte, Random walks and diffusion on networks,
  Phys. Rep. 716 (2017) 1--58.

\bibitem{ostrowski1973solutions}
A.~N. Ostrowski, Solutions of Equations in Euclidean and Banach Spaces,
  Academic Press, New York, 1973.

\bibitem{cucker1989alternate}
F.~Cucker, A.~G. Corbalan, An alternate proof of the continuity of the roots of
  a polynomial, Am. Math. Monthly 96 (1989) 342--345.

\bibitem{biggs1993algebraic}
N.~Biggs, Algebraic Graph Theory, 2nd Edition, Cambridge University Press,
  Cambridge, 1993.

\bibitem{godsil2013algebraic}
C.~Godsil, G.~F. Royle, Algebraic Graph Theory, Springer, Berlin, 2013.

\bibitem{tanaka2014random}
G.~Tanaka, C.~Urabe, K.~Aihara, Random and targeted interventions for epidemic
  control in metapopulation models, Sci. Rep. 4 (2014) 5522.

\bibitem{matsuki2019intervention}
A.~Matsuki, G.~Tanaka, Intervention threshold for epidemic control in
  susceptible-infected-recovered metapopulation models, Phys. Rev. E 100 (2019)
  022302.

\bibitem{gong2019modelling}
Y.~Gong, M.~Small, Modelling the effect of heterogeneous vaccination on
  metapopulation epidemic dynamics, Phys. Lett. A 383 (2019) 125996.

\bibitem{meloni2011modeling}
S.~Meloni, N.~Perra, A.~Arenas, S.~G{\'o}mez, Y.~Moreno, A.~Vespignani,
  Modeling human mobility responses to the large-scale spreading of infectious
  diseases, Sci. Rep. 1 (2011) 62.

\bibitem{wang2012safety}
B.~Wang, L.~Cao, H.~Suzuki, K.~Aihara, Safety-information-driven human mobility
  patterns with metapopulation epidemic dynamics, Sci. Rep. 2 (2012) 887.

\bibitem{nicolaides2013price}
C.~Nicolaides, L.~Cueto-Felgueroso, R.~Juanes, The price of anarchy in
  mobility-driven contagion dynamics, J. R. Soc. Interface 10 (2013) 20130495.

\bibitem{wang2017interplay}
B.~Wang, Y.~Han, G.~Tanaka, Interplay between epidemic spread and information
  propagation on metapopulation networks, J. Theor. Biol. 420 (2017) 18--25.

\bibitem{ferguson2006strategies}
N.~M. Ferguson, D.~A. Cummings, C.~Fraser, J.~C. Cajka, P.~C. Cooley, D.~S.
  Burke, Strategies for mitigating an influenza pandemic, Nature 442 (2006)
  448--452.

\bibitem{colizza2007modeling}
V.~Colizza, A.~Barrat, M.~Barthelemy, A.-J. Valleron, A.~Vespignani, Modeling
  the worldwide spread of pandemic influenza: baseline case and containment
  interventions, PLoS Med. 4 (2007) e13.

\bibitem{kitsak2010identification}
M.~Kitsak, L.~K. Gallos, S.~Havlin, F.~Liljeros, L.~Muchnik, H.~E. Stanley,
  H.~A. Makse, Identification of influential spreaders in complex networks,
  Nat. Phys. 6 (2010) 888--893.

\bibitem{ahajjam2018identification}
S.~Ahajjam, H.~Badir, Identification of influential spreaders in complex
  networks using hybridrank algorithm, Sci. Rep. 8 (2018) 11932.

\bibitem{nagatani2018metapopulation}
T.~Nagatani, G.~Ichinose, K.~Tainaka, Metapopulation model for
  rock--paper--scissors game: mutation affects paradoxical impacts, J. Theor.
  Biol. 450 (2018) 22--29.

\bibitem{kabir2019evolutionary}
K.~A. Kabir, J.~Tanimoto, Evolutionary vaccination game approach in
  metapopulation migration model with information spreading on different
  graphs, Chaos, Solitons \& Fractals 120 (2019) 41--55.

\bibitem{bonsall2004demographic}
M.~B. Bonsall, A.~Hastings, Demographic and environmental stochasticity in
  predator--prey metapopulation dynamics, J. Anim. Ecol. 73 (2004) 1043--1055.

\bibitem{cooper2012intermediate}
J.~K. Cooper, J.~Li, D.~J. Montagnes, Intermediate fragmentation per se
  provides stable predator-prey metapopulation dynamics, Ecol. Lett. 15 (2012)
  856--863.

\bibitem{tee2007eigenvectors}
G.~J. Tee, Eigenvectors of block circulant and alternating circulant matrices,
  New Zealand J. Math. 36 (2007) 195--211.

\bibitem{code}
Meng, L. Data sets and Python codes for epidemic dynamics on metapopulation
  networks under node2vec mobility. See
  \url{https://github.com/lingqime/Epidemic-dynamics-on-metapopulation-networks-with-node2vec-mobility}.
  Accessed on September 10, 2021.

\end{thebibliography}
\end{document}